\documentclass[aps,pre,superscriptaddress,nofootinbib,preprintnumbers]{revtex4-2}

\usepackage{graphicx}
\usepackage{graphics}
\usepackage{amsmath,amssymb}
\usepackage[usenames]{color}
\usepackage{subfigure}
\usepackage{hyperref}
\usepackage{breakurl}
\usepackage{enumitem}
\usepackage{lgreek}
\usepackage{enumitem}

\newcommand{\be}{\begin{equation}}
\newcommand{\ee}{\end{equation}}
\newcommand{\bea}{\begin{eqnarray}}
\newcommand{\eea}{\end{eqnarray}}
\newcommand{\ben}{\begin{enumerate}}
\newcommand{\een}{\end{enumerate}}
\newcommand{\bit}{\begin{itemize}}
\newcommand{\eit}{\end{itemize}}

\newcommand{\la}[1]{\label{#1}}

\newcommand{\Eq}[1]{Eq.~(\ref{#1})}

\newcommand{\Sec}[1]{Sec.~\ref{#1}}

\newcommand{\Fig}[1]{Fig.~\ref{#1}}

\def\nl{\nonumber \\}

\newcommand{\vv}[1]{\mathbf #1}							

\newcommand{\Del}{\boldsymbol \nabla}						
\newcommand{\bert}{\raise-0.45mm\hbox{\Large$\Box$}}			

\makeatletter
\newcommand*\bigcdot{\mathpalette\bigcdot@{.5}}
\newcommand*\bigcdot@[2]{\mathbin{\vcenter{\hbox{\scalebox{#2}{$\m@th#1\bullet$}}}}}
\makeatother

\definecolor{BrickRed}{cmyk}{0,0.89,0.94,0.28}					
\definecolor{MidnightBlue}{cmyk}{0.98,0.13,0,0.43}				
\definecolor{DarkGreen}{rgb}{0.100806,0.495968,0.209979}
\definecolor{orange}{rgb}{0.587167,0.354498,0.146197}

\begin{document}

\title{Leaking elastic capacitor as model for active matter}
	
\author{Robert Alicki}
\email{robert.alicki@ug.edu.pl}
\affiliation{International Centre for Theory of Quantum Technologies (ICTQT), University of Gda\'nsk, 80-308, Gda\'nsk, Poland}
\author{David Gelbwaser-Klimovsky}
\email{dgelbi@mit.edu}
\affiliation{Physics of Living Systems, Department of Physics, Massachusetts Institute of Technology, Cambridge, MA 02139, USA}
\author{Alejandro Jenkins}
\email{alejandro.jenkins@ucr.ac.cr}
\affiliation{International Centre for Theory of Quantum Technologies (ICTQT), University of Gda\'nsk, 80-308, Gda\'nsk, Poland}
\affiliation{Laboratorio de F\'isica Te\'orica y Computacional, Escuela de F\'isica, Universidad de Costa Rica, 11501-2060, San Jos\'e, Costa Rica}

\date{First version: 12 October, 2020.  This revision: 11 May, 2021.  To be published in Phys.\ Rev.\ E {\bf 103}}

\begin{abstract}
We introduce the ``leaking elastic capacitor'' (LEC) model, a nonconservative dynamical system that combines simple electrical and mechanical degrees of freedom.  We show that an LEC connected to an external voltage source can be destabilized (Hopf bifurcation) due to positive feedback between the mechanical separation of the plates and their electrical charging.  Numerical simulation finds regimes in which the LEC exhibits a limit cycle (regular self-oscillation) or strange attractors (chaos).  The LEC acts as an autonomous engine, cyclically performing work at the expense of the constant voltage source.  We show that this mechanical work can be used to pump current, generating an electromotive force without any time-varying magnetic flux and in a thermodynamically irreversible way.  We consider how this mechanism can sustain electromechanical waves propagating along flexible plates.  We argue that the LEC model can offer a qualitatively new and more realistic description of important properties of active systems with electrical double layers in condensed-matter physics, chemistry, and biology.
\end{abstract}


\maketitle


\section{Introduction}
\la{sec:intro}

\begin{figure} [t]
	\centering
	\includegraphics[height= 0.2 \textwidth]{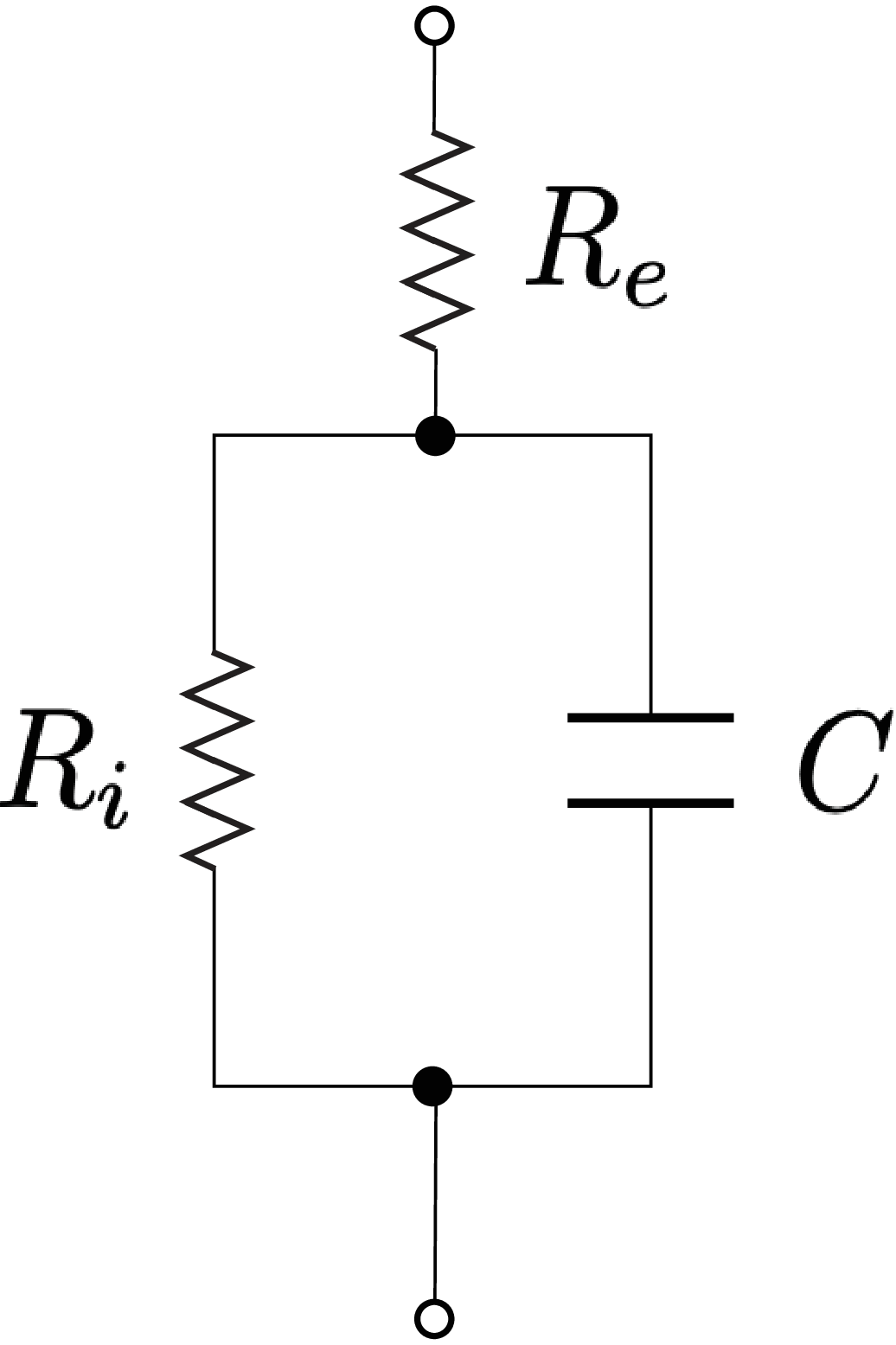}
\caption{\small Equivalent AC circuit for the electrical double layer, with capacitance $C$, internal resistance $R_i$, and external resistance $R_e$.  Adapted from Fig. 5-21 in \cite{Hamann}.\label{fig:equivalent}}
\end{figure}

The formation of an electrical double layer (EDL) at the interface between dissimilar materials is common to many systems of interest and has many practical applications.  Such EDLs are often treated in terms of an equivalent circuit with a capacitor and two resistors, one in parallel and the other in series with the capacitor, as shown in \Fig{fig:equivalent}.  This simple model can be applied to the electrode-electrolyte interface in batteries and fuel cells \cite{AL, Hamann}, $p$-$n$ junctions in photovoltaic and thermoelectric devices \cite{EvH}, cell membranes (based on lipid bilayers) in living organisms \cite{McL}, and double layers in inhomogeneous plasmas \cite{Raadu}.

The deformability of EDLs and the effect of deformations on their capacitance have long been a subject of interest in applied physics.  In 1966, Babakov, Ermishkin, and Lieberman reported that increasing the voltage applied to an artificial bimolecular lipid membrane caused it to deform in such a way that the capacitance increased, and they argued that such ``electromechanical properties'' could ``play an essential part in the activity of the cell membrane'' \cite{Babakov}.  These and other similar results \cite{Lauger, Rosen} motivated experimental and theoretical investigations into the interaction between the mechanical and the electrical properties of EDLs in biophysics \cite{White, Wobschall, Crowley} and later also in electrochemistry \cite{Feldman1986a, Feldman1986b, Feldman1987, Kornyshev, Partenskii1996}.

In \cite{Crowley}, Crowley introduced the term ``elastic capacitor'' (EC) to describe a model of the double layer in which the equilibrium configuration arises from a mechanical elasticity that counteracts the electrical attraction between the two charged plates.  For research connected to the EC in a technological context, see, e.g., \cite{Nieminen} and the treatment of ``memcapacitive systems'' in \cite{Pershin} (the latter reference is interesting in that it considers the dynamical response of the EC when subjected to external voltage pulses).  Partenskii and Jordan have shown how the EC model can be extended to account for certain properties of biological and chemical EDLs \cite{Partenskii2001, Partenskii2002, Partenskii2005, Partenskii2009}, and they have reviewed the subject from a pedagogical perspective in \cite{Partenskii2011}.

In this article we propose a new physical model for EDLs, which we call the {\it leaking elastic capacitor} (LEC).  It consists of a parallel plate capacitor in which the mechanical separation $X$ between the two plates is treated as a function of time in the relevant equations of motion.  Moreover, we take this $X(t)$ as determining not only the capacitance (as in the EC model), but also the internal conductivity and/or the potential applied to the capacitor by the external voltage source, as shown in \Fig{fig:LEC}.  Under certain conditions that we determine mathematically, this coupling of the mechanical $X(t)$ to the electrical degrees of freedom gives rise to a positive feedback that causes self-oscillation of $X$, the capacitor's charge $Q$, and the capacitor's voltage $V$.  The LEC therefore provides a remarkably simple model of an alternating-current (AC) generator.  In the framing of this model we have simplified the EDL structure as far as possible, abstracting all details about their implementations in a particular context, such as electrochemistry, solid state physics, plasmas, or biological membranes.

Self-oscillation can be defined as the generation and maintenance of a periodic motion, at the expense of a source of power that has no corresponding periodicity \cite{AVK}.  Self-oscillators are engines, which cyclically extract work from an underlying disequilibrium and whose operation is thermodynamically irreversible.  This physical approach to classical self-oscillators was pioneered by Le Corbeiller in \cite{LeC1} and \cite{LeC2}, and has more recently been advocated and explored in \cite{SO}.  This work extraction is manifested as an {\it active, nonconservative force} upon a macroscopic degree of freedom that serves as the engine's piston or turbine.  In the literature, such a nonconservative force has been introduced as a negative friction or anti-damping (as in the van der Pol model \cite{vdP1, vdP2} and in treatments of self-propelled Brownian particles \cite{active-brownian, self-propelled}), as a circulatory force (see, e.g., the treatment of the nonconservative force in stochastic thermodynamics \cite{Seifert}), as a time delay in the oscillator's response (see, e.g., \cite{Minorsky} and \cite{bio-oscillators}), or as an external periodic forcing (see, e.g., the model of Brownian motors in \cite{Hanggi}).  None of these mathematical descriptions, however, provide a physically realistic account of how the nonconservative force is generated dynamically by the autonomous operation of the engine.

In practice, the active, nonconservative force driving the self-oscillation requires a positive feedback between at least two macroscopic degrees of freedom and can arise only in open systems coupled to an external thermal or chemical disequilibrium.  In the case of the LEC, this disequilibrium corresponds to the external voltage source.  Self-oscillation results from a positive feedback between the plate separation $X$ and the capacitor charge $Q$.  The necessary nonconservation is provided by the Ohmic conductivity (i.e., by the leakiness).

We also show that the work extracted cyclically by the LEC from the external voltage can be used to generate an electromotive force (emf), which may pump charges to build up potential differences or to drive the electric charges along a closed path.  On the emf as the integral of an active nonconservative force, and on the impossibility of accounting for it using only potentials, see \cite{emf}.  Note that the transformation of the external $V_0$ into cyclical work (from which an emf can be obtained) is an irreversible process and that some of the free energy corresponding to the external voltage source $V_0$ is necessarily dissipated by the ohmic conductivity.

With minor generalizations, the equations of motion that we derive for the LEC can be applied to a very broad class of active systems.  (See the Appendix for the precise sense in which we are here using the term {\it active}.)  These include such diverse examples as the ``beating mercury heart'' \cite{mercury}, the electron shuttle \cite{shuttle1, shuttle2}, and proposed dynamical models for the operation of photovoltaic \cite{solarcells, AGJ}, thermoelectric \cite{thermocells}, fuel \cite{fuelcells}, and electrochemical cells (i.e., batteries) \cite{battery}.  Double layers are present in all of these systems.  For photovoltaic cells, collective oscillations in the terahertz frequency range have already been reported. \cite{Nakanishi2012, Guzelturk2018}.\footnote{ A very close analog of the LEC applies also to simple steam engines (see the discussion of the ``putt-putt engine'' in \cite{SO} and \cite{AGJ}), even though there the relevant interface is not electrically charged.}

There is now ample evidence of the importance of the mechanical properties of lipid bilayers to the functionality of biological cell membranes \cite{membrane-forces}.  Passive and active ion channels can be described in an LEC model in terms of controlled resistance and emf, respectively (see, e.g., \cite{AK}).  A description based on work extraction and charge pumping by a self-oscillating LEC may be both simpler and more realistic than the Hodgkin-Huxley \cite{Hodgkin-Huxley}, the FitzHugh-Nagumo \cite{FitzHugh, Nagumo}, and other models of excitable cell membranes (on the active dynamics of these models, see, e.g., \cite{NeuronDynamics} and references therein).  Moreover, if the rigid plates are replaced with flexible surfaces, the system of ordinary differential equations for the LEC is replaced by coupled wave and reaction-diffusion equations with solutions corresponding to self-sustained traveling waves.  In the last section of this paper we argue that this offers a promising new model for neural signaling and electric energy transport in biological systems.

It is also worth noting here that Alfv\'en and others have stressed the importance of EDLs in inhomogeneous plasmas and their role in particle acceleration in the laboratory and in astrophysics.  Such phenomena are difficult to account for using the equations of magnetohydrodynamics, leading Alfv\'en to advocate instead a phenomenological approach based on simple circuit models \cite{Alfven}.  Experiments with gaseous plasmas have found coherent EDL oscillations, with the fast component at around the plasma frequency \cite{Torven, Volwerk}.  Amusingly, in \cite{Alfven}, Alfv\'en also suggested an analogy between plasma double layers and biological membranes.  In biophysics, Amin proposed in 1982 that EDLs might play an important role in the active transport of water in plants \cite{Amin}, a hypothesis that has since found some theoretical and experimental support \cite{Zimmermann}.  Plants generate audio and ultrasound signals that may be associated with such transport \cite{Perelman, Gagliano}.  The range of active systems to which the LEC model could potentially be relevant is therefore very broad.

This is an interdisciplinary theoretical investigation.  The fundamental problem addressed pertains to nonequilibrium thermodynamics: to understand how an active EDL can extract work in a sustained and thermodynamically irreversible way.  The simple electromechanical LEC model is presented in \Sec{sec:model}.  In \Sec{sec:SO} we apply to that model certain concepts and techniques from the mathematical theory of dynamical systems, in order to determine the conditions in which the LEC can self-oscillate.  In \Sec{sec:numerical} we describe the results of some numerical simulations of a particular nonlinear implementation of the LEC, which show that the LEC is capable of both periodic self-oscillation and chaotic behavior.  In \Sec{sec:emf} we show, in terms of classical electrodynamics, that the LEC can generate an emf despite the absence of any time-varying magnetic flux.  This allows us to connect our results to the understanding of active matter in electrochemistry and biophysics.  In \Sec{sec:traveling} we consider how to extend the LEC model to accommodate self-sustained traveling waves.  The mathematical treatment offered here is neither generic nor exhaustive, while the electromechanical models considered are the simplest ones, consistent with basic thermodynamic principles and other relevant physical constraints, that allow for active EDL dynamics.  Many details, particularly on the traveling wave model in \Sec{sec:traveling}, are left for future investigation.


\section{LEC model}
\la{sec:model}

\begin{figure}[t]
	\centering
\includegraphics[height=0.2 \textwidth]{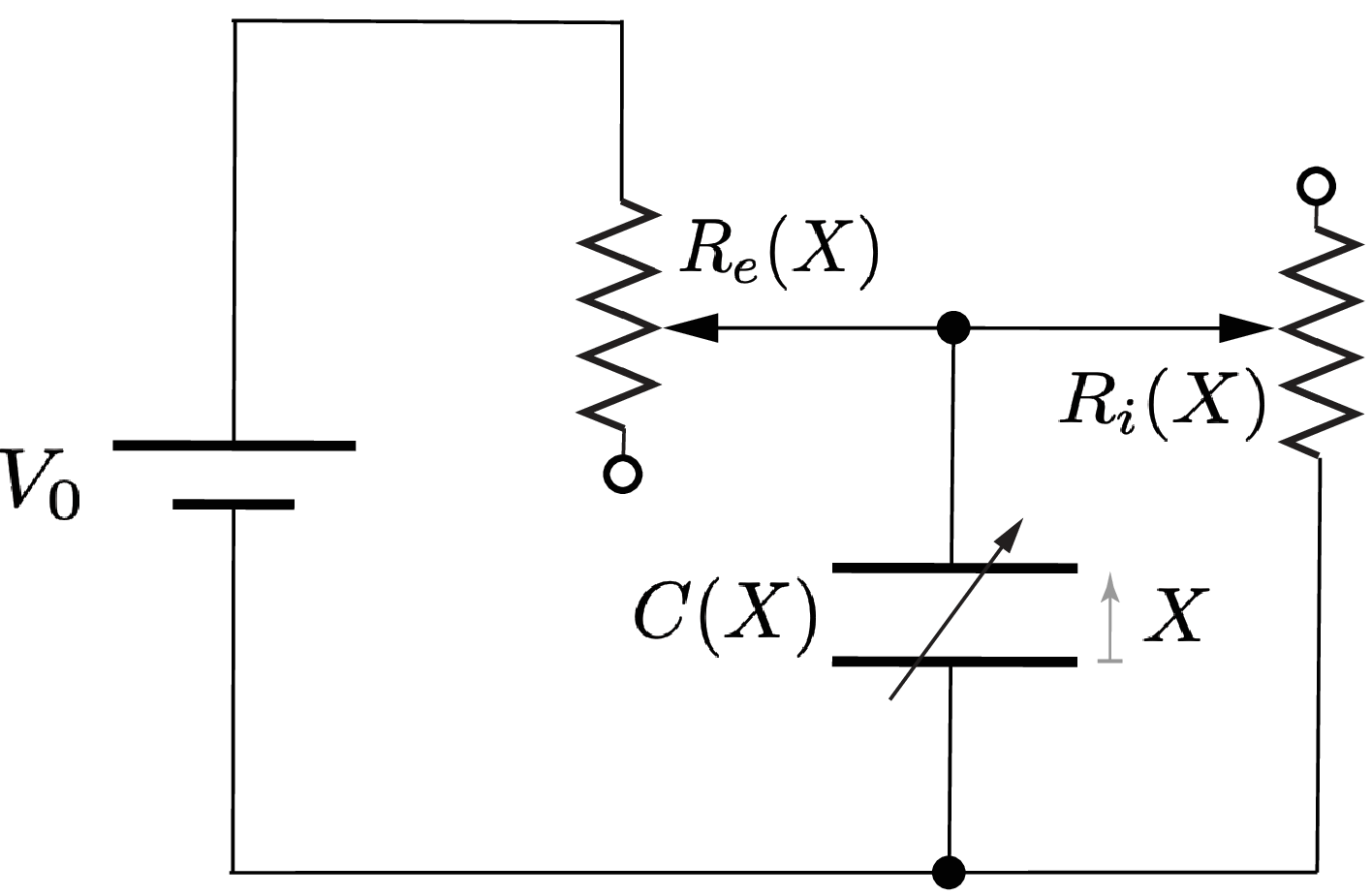}
\caption{The leaking elastic capacitor (LEC) model: an electromechanical circuit containing a constant source of voltage $V_0$, a capacitor $C$ with variable plate separation $X$, and two resistors $R_e$ and $R_i$. The values of the capacitance and of the two resistances depend on $X$.\label{fig:LEC}}
\end{figure} 

Consider the LEC described by the electro-mechanical circuit shown in \Fig{fig:LEC}, which consists of a capacitance $C(X)$, an internal resistance $R_i (X)$  (``leakage''), and an external resistance $R_e(X)$ that can control the potential supplied to the LEC by the constant external voltage $V_0$.  The mechanical degree of freedom is described by the dynamical variable $X(t)$, representing the distance between the two parallel plates of the capacitor.  The second dynamical variable is the charge $Q(t)$ accumulated in the capacitor.  By Kirchhoff's current law,
\be
\dot Q = - \Gamma(X) Q + I(X) ,
\la{eq:Qlaw}
\ee
where
\be
\Gamma(X) = \left[ \frac{1}{C(X) R_i(X)} + \frac{1}{C(X) R_e(X)} \right] > 0 , \quad I(X) = \frac{V_0}{R_e(X)} .
\label{eq:Gamma}
\ee
One could consider a more general circuit in which the external voltage $V_0$ is also controlled by the dynamical variable $X$.  Since the calculations presented below use an arbitrary function $I(X)$, they are also valid in this more general case.  One might also easily extend the model to have $\Gamma$ and $I$ depend on both $X$ and $Q$, but for simplicity we do not do that in this article.

The mechanical equation of motion (Newton's law) for $X$ is
\be
\ddot X + \gamma \dot X + \frac 1 M \frac{\partial}{\partial X} U(X,Q) = 0 , 
\la{eq:Xlaw}
\ee
where $M$ is the mass of the moving plate and $\gamma > 0 $ is the damping coefficient. The potential $U$ can be decomposed into an electrostatic and a mechanical part:
\be
U(X,Q) = \frac{Q^2}{2 C(X)} + U_m (X) .
\label{eq:U}
\ee
The mechanical potential $U_m$ should include a short-range repulsion to prevent an unphysical crossing of the capacitor's plates.\footnote{Some electrochemical models allow for negative capacitance in the vicinity of $Q = 0$, which in the simple EC picture would correspond to $X < 0$.  Such negative capacitance might result from the non monotonous distribution of ionic density associated with ``super-equivalent'' adsorption,  ``overscreening'', or ``overcharging'' \cite{Partenskii2001}.  We do not consider such possibilities here.}

The dependence of $\Gamma$ or/and $I$ on $X$, together with the dependence of $U$ on $Q$, provide the feedback necessary for the emergence of self-oscillations.  Equations \eqref{eq:Qlaw} and \eqref{eq:Xlaw} describe an autonomous dynamical system, which can be expressed in terms of three coupled first-order differential equations for the dependent variables $X$, $P = \dot{X}$, and $Q$:
\be
\left\{ \begin{array}{l}
\dot X = P , \\
\dot P = -\gamma P +  f(X,Q) , \\
\dot Q = - \Gamma(X)  Q + I(X) ,
\end{array} \right.
\la{eq:dyn}
\ee
with a force per unit mass 
\be
f(X, Q) = -\frac 1 M \frac{\partial}{\partial X} U(X, Q) .
\la{eq:force}
\ee

For any physical system, the damping rates $\gamma$ and $\Gamma(X)$ must be positive.  In the self-oscillating regime, \hbox{$M f = - \partial U / \partial X$} is the active, nonconservative force that cyclically extracts work from the external $V_0$.  The infinitesimal work
\be
\delta W = M f(X, Q) \, dX
\la{eq:dW}
\ee
is an inexact differential due to the time-dependence of $Q$ introduced by \Eq{eq:Qlaw}.  Note that \Eq{eq:Qlaw} represents a thermodynamically irreversible dynamics, since it involves Ohmic dissipation.  The fact that engines capable of generating positive power must operate irreversibly has recently been stressed, e.g., in \cite{Ouerdane} and \cite{dissipation-induced}.


\section{Self-oscillation}
\la{sec:SO}

To establish that the dynamical system of \Eq{eq:dyn} admits a self-oscillatory regime, we study the stability of the equilibrium configuration (fixed point) $\{X_0,P_0 ,Q_0\}$ for \Eq{eq:dyn}.  If, as the parameters of the system are varied, this equilibrium goes from stable to unstable, a Hopf bifurcation occurs, leading to self-oscillation \cite{SO}.

The equilibrium is determined by the equations
\be
P_0 = 0, \quad f(X_0 , Q_0) = 0, \quad Q_0 = \frac {I(X_0) }{\Gamma(X_0)} .
\la{eq:fixed}
\ee
Introducing the dimensionless variables
\be
x = \frac{X - X_0}{X_0}, \quad q =\frac{Q - Q_0}{Q_0}
\la{eq:xq}
\ee
and a normalized velocity $p = \dot x$ with the dimension of time$^{-1}$, one obtains a set of evolution equations for the deviation from equilibrium. For small deviations, nonlinear terms can be neglected and \Eq{eq:dyn} reduces, in matrix form, to
\be
\begin{pmatrix} \dot x  \\ \dot p \\ \dot q \end{pmatrix} = 
\begin{pmatrix}
0 & 1 & 0 \\
-\Omega_0^2 & - \gamma &- a \\
b & 0 & -\Gamma_0 
\end{pmatrix}
\begin{pmatrix} x \\ p \\ q \end{pmatrix} .
\la{eq:lin}
\ee
These matrix elements contain the parameters
\bea
\Omega_0^2 &=&- \frac{\partial}{\partial X} f(X_0, Q_0), \quad \Gamma_0 =  \Gamma(X_0) , \nl
a &=& - \frac{Q_0}{X_0}  \frac{\partial}{\partial Q} f(X_0, Q_0), \quad
b = X_0 \left[ Q_0^{-1} \frac{d I}{dX}(X_0) - \frac{d\Gamma}{dX}(X_0) \right] .
\la{eq:params}
\eea

Stability of the fixed point is determined by the eigenvalues of the $3\times 3$ matrix in \Eq{eq:lin}, given by the roots $\lambda_0 , \lambda_+$, and $\lambda_-$ of the characteristic equation
\be
(\lambda + \Gamma_0) \left[ \lambda (\lambda + \gamma) +\Omega_0^2 \right] + ab = 0 .
\la{eq:eigen}
\ee
Simple approximate formulas for $\lambda_j$ can be obtained in the weak damping regime (i.e.  $\gamma, \Gamma_0 \ll \Omega_0$), as long as the product $ab$ (which characterizes the strength of the feedback) is also small. More precisely, since for typical choices of mechanical potential $U_m(X)$ one obtains $a \simeq \Omega_0^2$ [see \Eq{eq:ab} below], we assume that $|b| \ll \Omega_0$ and then obtain that
\bea
\lambda_0 &\simeq& - \left( \Gamma_0 + \frac{ab}{\Omega_0^2} \right) , \nl
\lambda_{\pm} & \simeq& \pm i \Omega_0 + \frac{1}{2}  \left( \frac{ab}{\Omega_0^2} - \gamma \right)
\la{eq:eigen1}
\eea
satisfy \Eq{eq:eigen} up to small corrections.

For large enough $ab$, the real part of $\lambda_{\pm}$ in \Eq{eq:eigen1} is positive, which implies that the amplitude of small oscillations about equilibrium, with angular frequency $\Omega_0$, increases exponentially with time until they are limited by nonlinearities.  This corresponds to a Hopf bifurcation \cite{Strogatz}.  In \Sec{sec:numerical} we show numerically how the nonlinear effects stabilize the amplitude of the oscillations, leading either to a limit cycle or to a strange attractor.

The bifurcation can be found without the weak-damping approximation by using the test functions
\be
x(t) = x_0 \cos \omega t , \quad  q(t) = q_0 \sin (\omega t + \alpha) .
\la{eq:test}
\ee
The solution with real $\omega$ in \Eq{eq:test} corresponds to the critical condition for which the oscillations show zero linear damping, which is the Hopf bifurcation point.  Inserting \Eq{eq:test} into \Eq{eq:lin} we obtain
\bea
- \omega^2 x_0 \cos \omega t &=& - \Omega_0^2 x_0 \cos \omega t + \gamma \omega x_0 \sin \omega t - a q_0 \sin (\omega t + \alpha) \la{eq:test1} , \\
\omega q_0 \cos (\omega t + \alpha) &=& b x_0 \cos \omega t - \Gamma_0 q_0 \sin (\omega t + \alpha) . \la{eq:test2}
\eea
Evaluating \Eq{eq:test1} at $\omega t = \pi /2$ we obtain
\be
\cos \alpha = \frac{\gamma \omega x_0}{a q_0} .
\la{eq:cosa}
\ee
Evaluating \Eq{eq:test2} at $\omega t = \pi /2$ and substituting \Eq{eq:cosa} we get
\be
- \omega q_0 \sin \alpha = - \Gamma_0 q_0 \cos \alpha = - \frac{\Gamma_0 \gamma \omega x_0}{a} .
\la{eq:sina}
\ee
Evaluating \Eq{eq:test1} at $t=0$ and combining the result with \Eq{eq:sina} we arrive at
\be
\omega^2 = \Omega_0^2 + \gamma \Gamma_0 .
\la{eq:omega1}
\ee

Meanwhile, \Eq{eq:test1} can be rewritten as
\be
-q_0 \sin (\omega t + \alpha) = \frac{ \left(\Omega_0^2 - \omega^2 \right) x_0 \cos \omega t - \gamma \omega x_0 \sin \omega t}{a} = \frac{ - \Gamma_0 \gamma x_0 \cos \omega t - \gamma \omega x_0 \sin \omega t}{a} ~,
\la{eq:test1-1}
\ee
where we have used \Eq{eq:omega1} in the last step.  Substituting \Eq{eq:test1-1} in \Eq{eq:test2} and evaluating the result at $t=0$ gives
\be
\frac{a \omega q_0}{\gamma x_0} \cos \alpha = \frac{ab}{\gamma} - \Gamma_0^2 .
\la{eq:cosa-1}
\ee
Substituting \Eq{eq:cosa-1} into \Eq{eq:cosa} we arrive at
\be
\omega^2 = \frac{ab}{\gamma} - \Gamma_0^2 .
\la{eq:omega2}
\ee
In light of the results of Eqs.\ \eqref{eq:omega1} and \eqref{eq:omega2} we introduce the feedback parameter
\be
\eta = \frac{ab}{\gamma \cdot \left( \Omega_0^2 +  \Gamma_0^2 + \gamma\Gamma_0 \right)} ,
\la{eq:eta}
\ee
such that $\eta = 1$ corresponds to the Hopf bifurcation.  It is easy to see, by comparing this to the stability analysis of Eqs.\ \eqref{eq:eigen} and \eqref{eq:eigen1} (or to the numerical results of the nonlinear equations of motion in \Sec{sec:numerical}), that $\eta < 1$ corresponds to a stable fixed point and $\eta > 1$ to the self-oscillation.  We can interpret this as a condition on the strength of the positive feedback parameter $ab$.

Note that the results of Eqs.\ \eqref{eq:omega1} and \eqref{eq:omega2} establish that, if the feedback $ab$ is large enough, then the LEC can self-oscillate in the strongly overdamped regime ($\gamma \gg \Omega_0$) with a frequency $\omega$ high compared to the resonant $\Omega_0$.  In that case, $\omega$ is controlled by the damping rates.  This corresponds to the highly nonsinusoidal ``relaxation oscillation'' regime (see, e.g., \cite{SO}).  Physically speaking, relaxation oscillations are characterized by the fact that no significant energy is stored by the oscillator from one period of the limit cycle to the next.  Instead, the external source must supply almost all of the oscillator's energy once per period \cite{Groszkowski}.  We expect that this relaxation-oscillation regime should be common in active soft-matter systems, which are strongly dissipative (see also the discussion of self-sustained traveling waves in \Sec{sec:traveling}).

\begin{figure}[t]
\centering
	\subfigure[]{\includegraphics[height=0.2 \textwidth]{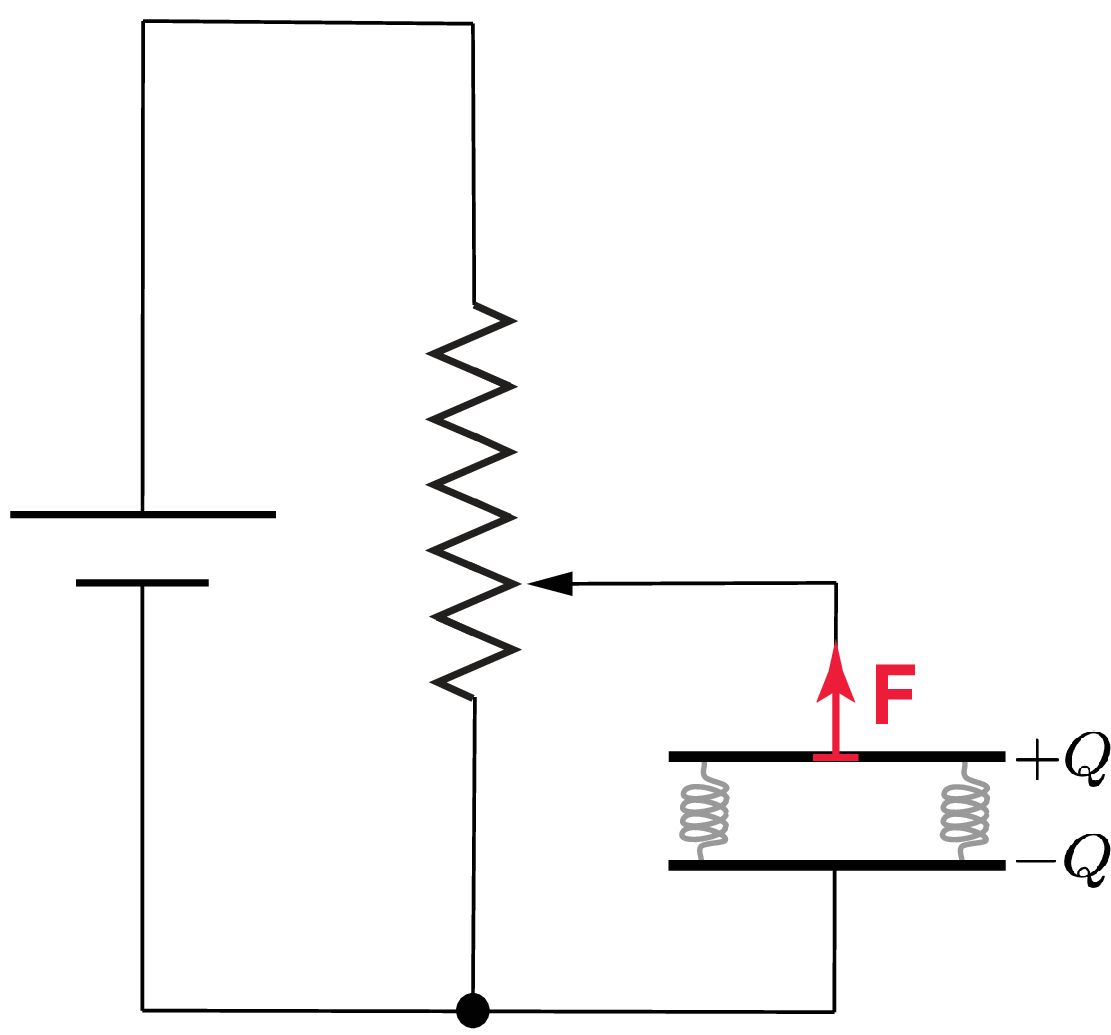}} \hskip 1.5 cm
	\subfigure[]{\includegraphics[height=0.2 \textwidth]{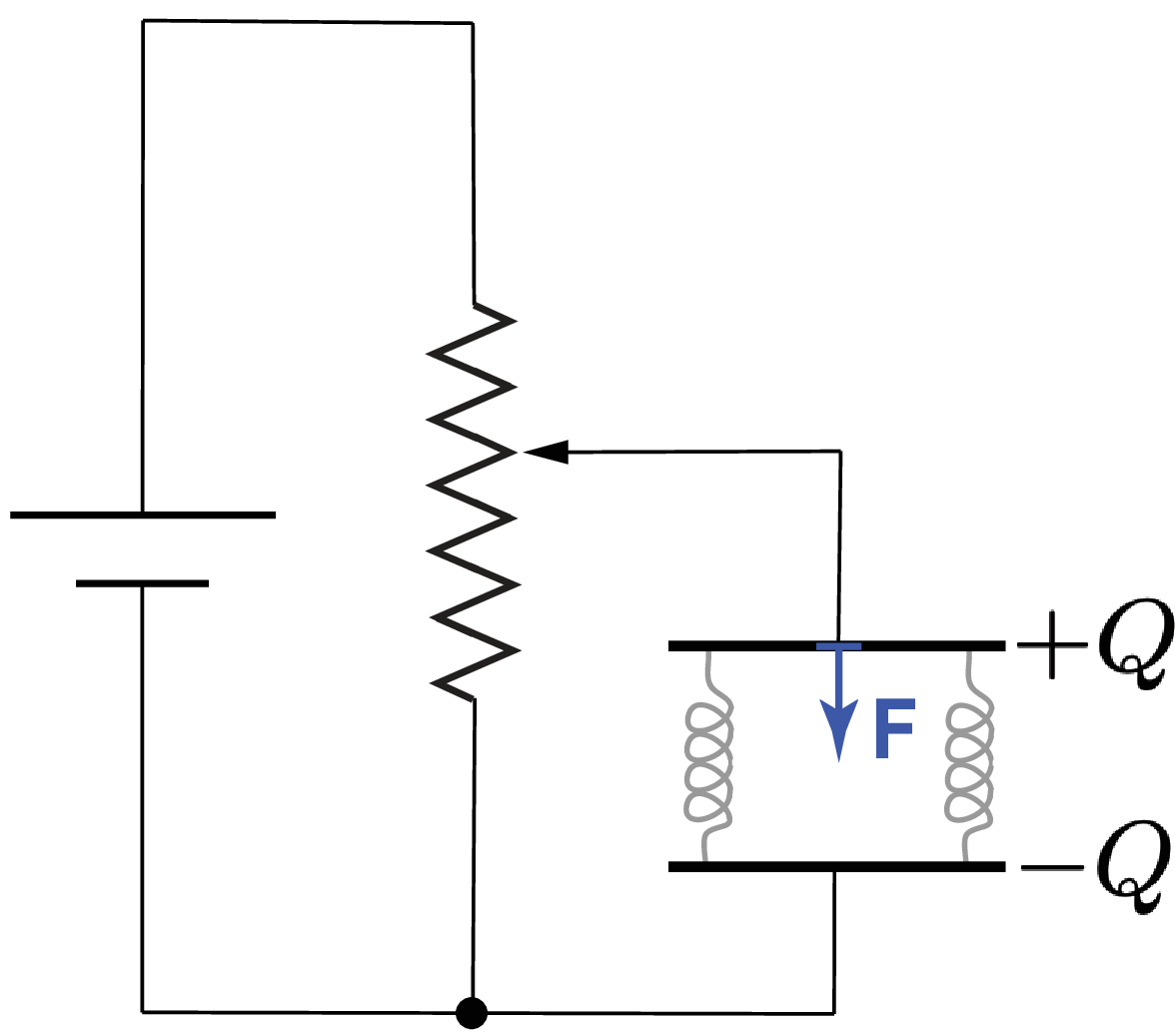}}
\caption{(a) Self-oscillating LEC expanding ($\dot X > 0$) as the elastic force exceeds the electrostatic attraction, so that the net force $F$ tends to push the plates apart.  (b) Self-oscillating LEC contracting ($\dot X < 0$) as an increased electrostatic attraction results in a net $F$ that pulls the plates together.  Note that, in the particular circuits shown here, $X$ acts a voltage divider because $R_i (X) + R_e (X) = $ const.  Therefore, $X$ directly modulates the potential difference between the plates.\label{fig:LEC-ab}}
\end{figure}

That the plates of the capacitor in \Fig{fig:LEC} may self-oscillate due to a positive feedback between $X$ and the charge $Q$ of the capacitor is easy to understand qualitatively.  For simplicity and definiteness, consider the particular example of an LEC in which the sum $R_i (X) + R_e (X)$ is fixed and $X$ therefore acts as a simple voltage divider, as show in \Fig{fig:LEC-ab}.  A greater $Q$ increases the electrostatic attraction between the plates, and therefore tends to reduce $X$.  A smaller $X$ reduces the voltage applied to the plates and therefore also the charge $Q$.  This, in turn, reduces the electrostatic attraction between the plates.  Mechanical elasticity can then cause the separation $X$ to increase, as shown in \Fig{fig:LEC-ab}(a).  This favors the charging of the plates and therefore tends to pull them back together, as in \Fig{fig:LEC-ab}(b).  If the electrostatic force (proportional to $-Q^2$) varies in phase with $\dot X$, then the motion of the plates will be effectively anti-damped, leading to an electromechanical self-oscillation powered by the external voltage source $V_0$.


\section{Numerical simulations}
\la{sec:numerical}

To illustrate the qualitative features of the dynamics of the LEC as an engine, we perform numerical simulations for a particular nonlinear implementation.   For an ideal parallel plate capacitor, the capacitance can be expressed as
\be
C(X) \equiv C(x) = \frac{C_0}{1 + x} ,
\la{cap1}
\ee
in terms of the dimensionless $x$ of \Eq{eq:xq}, where $C_0$ is the equilibrium capacitance [$C_0 = C(X_0)$].  In order to reduce the number of free parameters, henceforth we take a constant external resistance $R_e(X) = R_0$.  As a further simplification, we consider the internal resistance as a switch that closes the circuit when the capacitor is squeezed and opens it when the capacitor expands.  A simple way to implement this is to take
\be
R_i (X) \equiv R_i (x) = R_0 e^{\beta x} ,
\la{eq:res1}
\ee
for a dimensionless $\beta > 0$.  Then
\be
\Gamma(X) \equiv \Gamma(x) = \frac 1 2 \Gamma_0 (1+x) \left( 1 + e^{-\beta x} \right), \quad  \Gamma_0 =\frac{2}{ C_0 R_0} ,
\la{eq:Gammax}
\ee
where $2\Gamma_0^{-1}= R_0 C_0 $ is the time constant for the RC circuit.

Given the aims of our present investigation, we do not attempt to derive the mechanical potential $U_m (X)$ from a microphysical picture of a particular implementation of an EDL, such as the electrochemical Gouy-Chapman model \cite{Gouy}.  This is a task that we leave for future investigation.  For our simulations we simply use
\be
U_m (X) =  M \frac \sigma X , \quad  \sigma > 0 ,
\la{eq:pot1}
\ee
which prevents the crossing of the plates ($X < 0$), as discussed in \Sec{sec:model}, and is qualitatively consistent with the picture of a pressure exerted by a gas of ions confined within the EDL.  The phenomenological parameter $\sigma$ in \Eq{eq:pot1} can be taken from the measured values of $X_0$ and $Q_0$ at equilibrium.  The property that $X_0 \to \infty$ for $Q_0 \to 0$ is consistent with the electrochemical double layer dissolving if the electrode is uncharged.\footnote{Note that the elasticity of the LEC (represented by the springs drawn in \Fig{fig:LEC-ab}) must, for the choice of $U_m$ in \Eq{eq:pot1}, be interpreted as the first-order term in the power series in $X$ of the force $M f(X,Q)$ in \Eq{eq:force}, expanded about the fixed point $\{X_0, Q_0\}$.  Thus, the ``elasticity'' of the LEC here depends on the charge and is not constant over time.}  On the other hand, for a bimolecular lipid membrane $X_0$ remains finite for $Q_0 \to 0$, so that an elastic term should be added to the potential of \Eq{eq:pot1}.  It is important to stress that our conclusions do not depend in any significant way on the details of the nonlinear interaction corresponding to the choice of $U_m$.

Using the parametrization described above and inserting the equilibrium conditions of \Eq{eq:fixed}, we obtain equations of motion of the form
\bea
\dot q &=& -\frac{1}{2}\Gamma_0 \left[(1+x)(1+q) \left(1+ e^{-\beta x}  \right) -2 \right] , \nl
\ddot x &=& - \gamma \dot x - \frac{1}{2}\Omega_0^2 \left[(1 + q)^2 - \frac{1}{(1 + x)^2} \right] ,
\la{eq:motion1}
\eea
where 
\be
\Omega_0^2 = \frac{Q_0^2}{M C_0 X_0^2} .
\la{eq:Omega}
\ee

By choosing the units of time appropriately, we can always set $\Omega_0 = 1$.  We can then plot the trajectories projected on the $(x,q)$ plane for various values of the (now dimensionless) parameters $\gamma , \Gamma_0$, and $\beta$, as shown in \Fig{fig:trajectories}. To determine the feedback parameter of \Eq{eq:eta} we compute first
\be
a = \Omega_0^2 , \quad b = \frac{1}{2}\Gamma_0 (\beta -2),
\la{eq:ab}
\ee
so that
\be
\eta = \frac{\Gamma_0 (\beta - 2)}{2 \gamma (1 + \Gamma_0^2 + \gamma\Gamma_0)} .
\label{eq:eta1}
\ee

The experimentally accessible variable is usually the capacitor voltage $V_c (t) = Q(t) /C(X(t))$.  A plot of its normalized, dimensionless counterpart
\be
v(t) = \frac{V_c(t) - V_c^{(0)}}{V_c^{(0)}} =  x(t) + q(t) + x(t)q(t) , \quad V_c^{(0)} = \frac{Q_0}{C_0} ,
\la{eq:voltage}
\ee
is given in \Fig{fig:trajectories} for various choices of $\gamma , \Gamma_0$, and $\beta$.

A useful mathematical object is a power spectrum of the voltage $v(t)$, defined as 
\be
S(\omega) = \lim_{T\to\infty} S_T(\omega), \quad S_T(\omega) = \frac{1}{2\pi T} \left| \int_0^T   v(t) e^{-i\omega t} dt \right|^2 .
\la{eq:pspectrum}
\ee
This can be applied to distinguish the limit-cycle from the strange-attractor regimes, as illustrated in \Fig{fig:pspectrum}.  On the use of this power spectrum for studying the chaotic regime of a dynamical system, see, e.g., \cite{powerspec}.

\begin{figure} [t]
\centering
	\subfigure[]{\includegraphics[height=0.45 \textwidth]{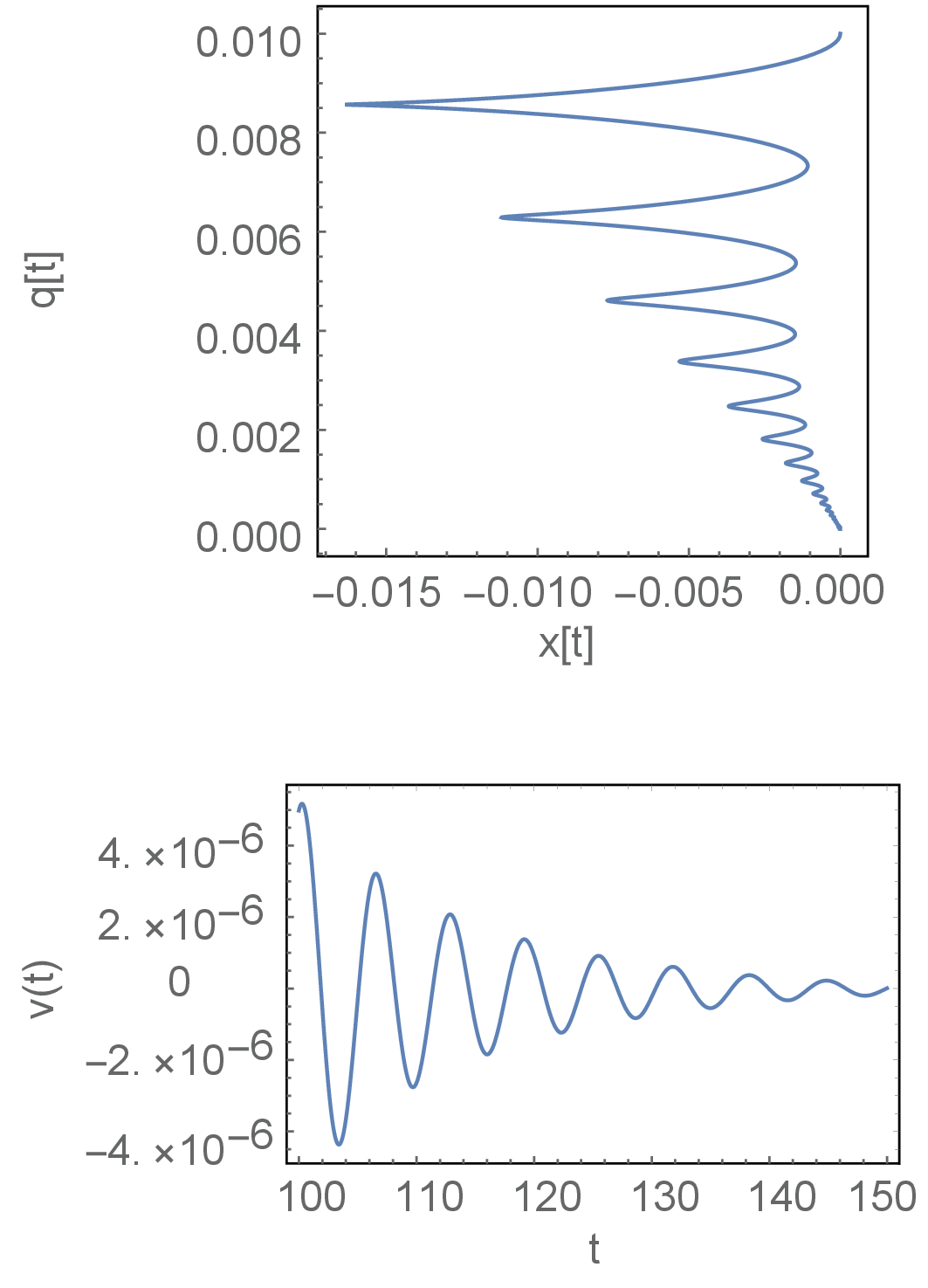}} \hskip 2 cm
	\subfigure[]{\includegraphics[height=0.45 \textwidth]{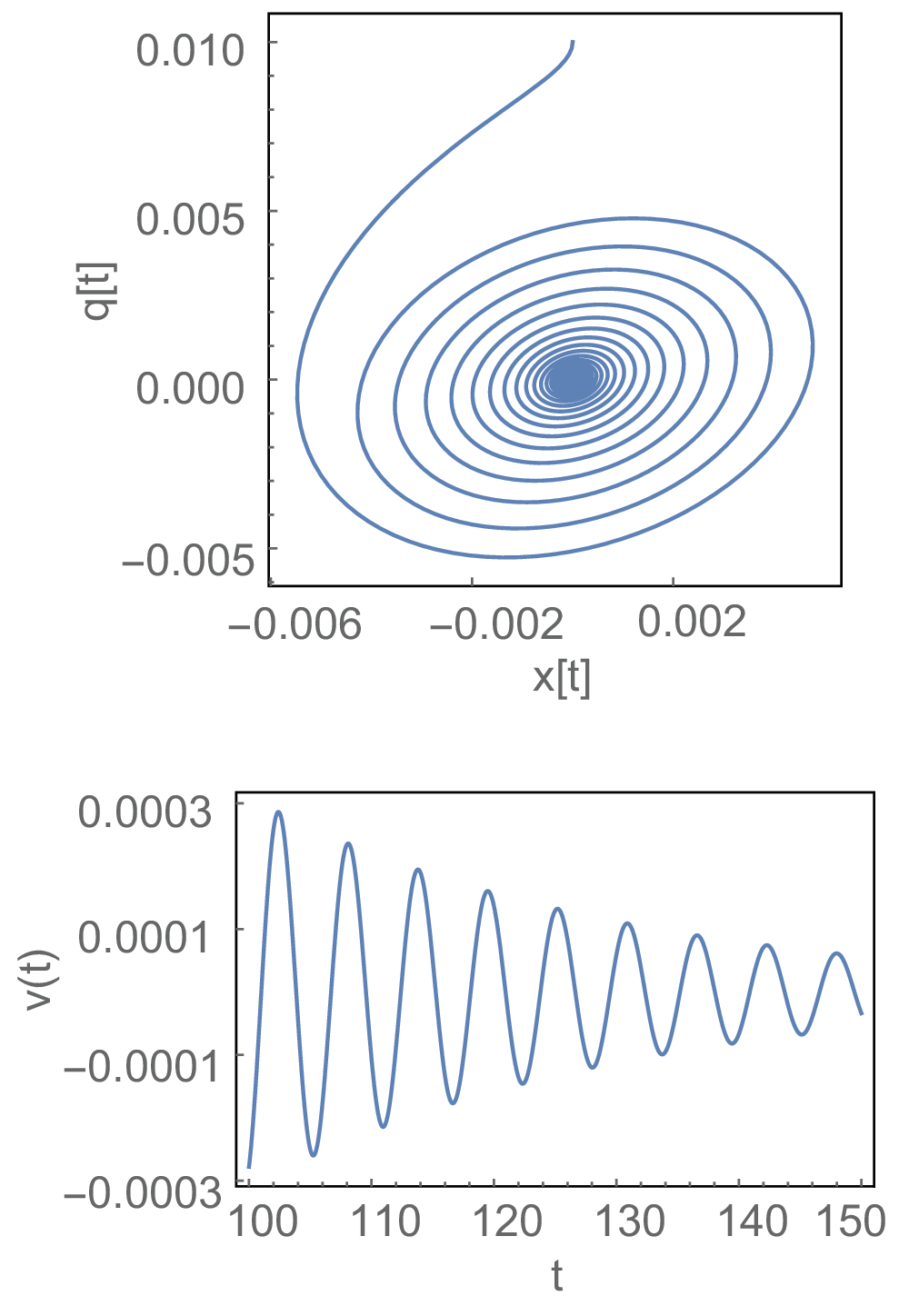}} \\ \vskip 0.7 cm
	\subfigure[]{\includegraphics[height=0.45 \textwidth]{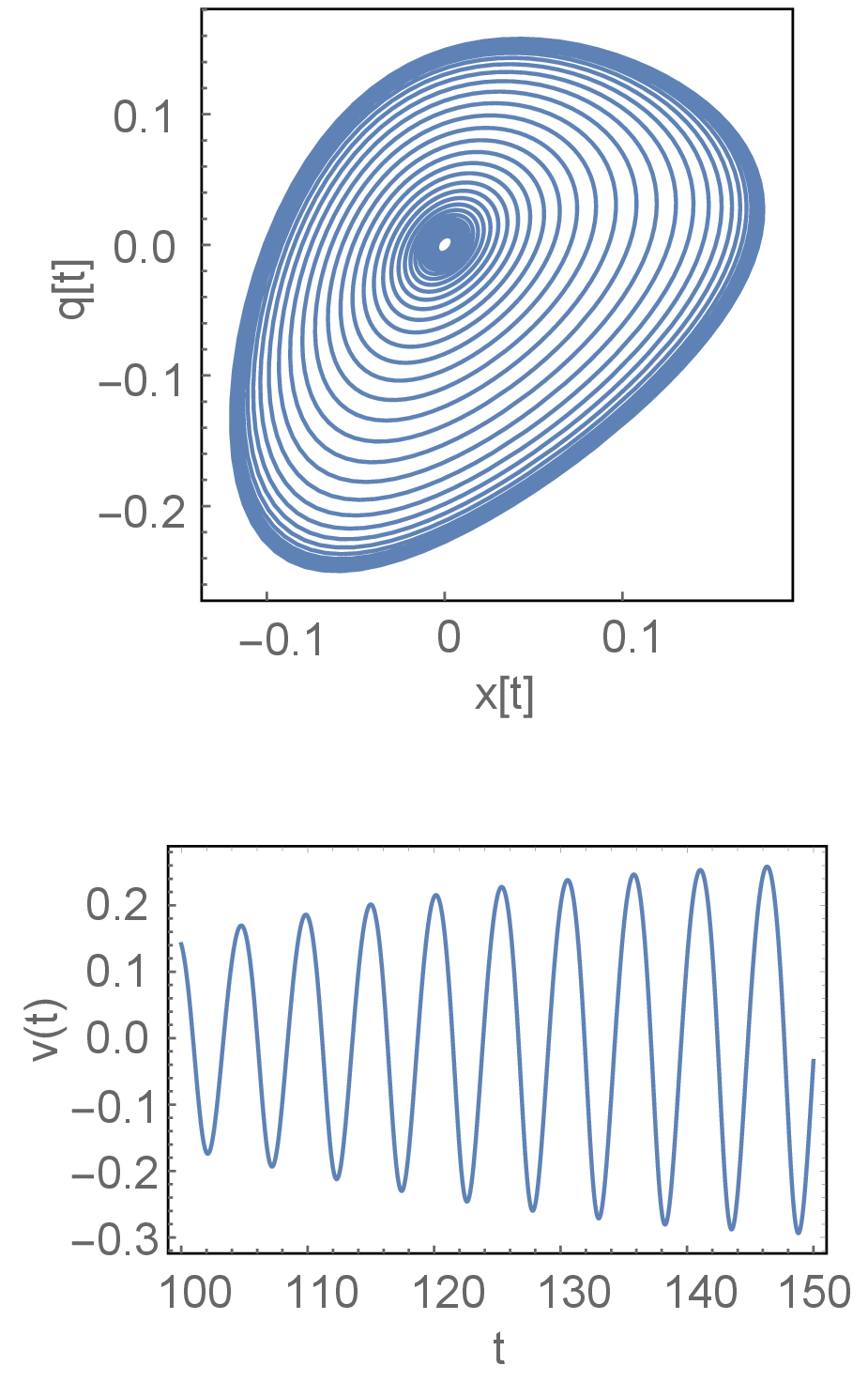}} \hskip 2 cm
	\subfigure[]{\includegraphics[height=0.45 \textwidth]{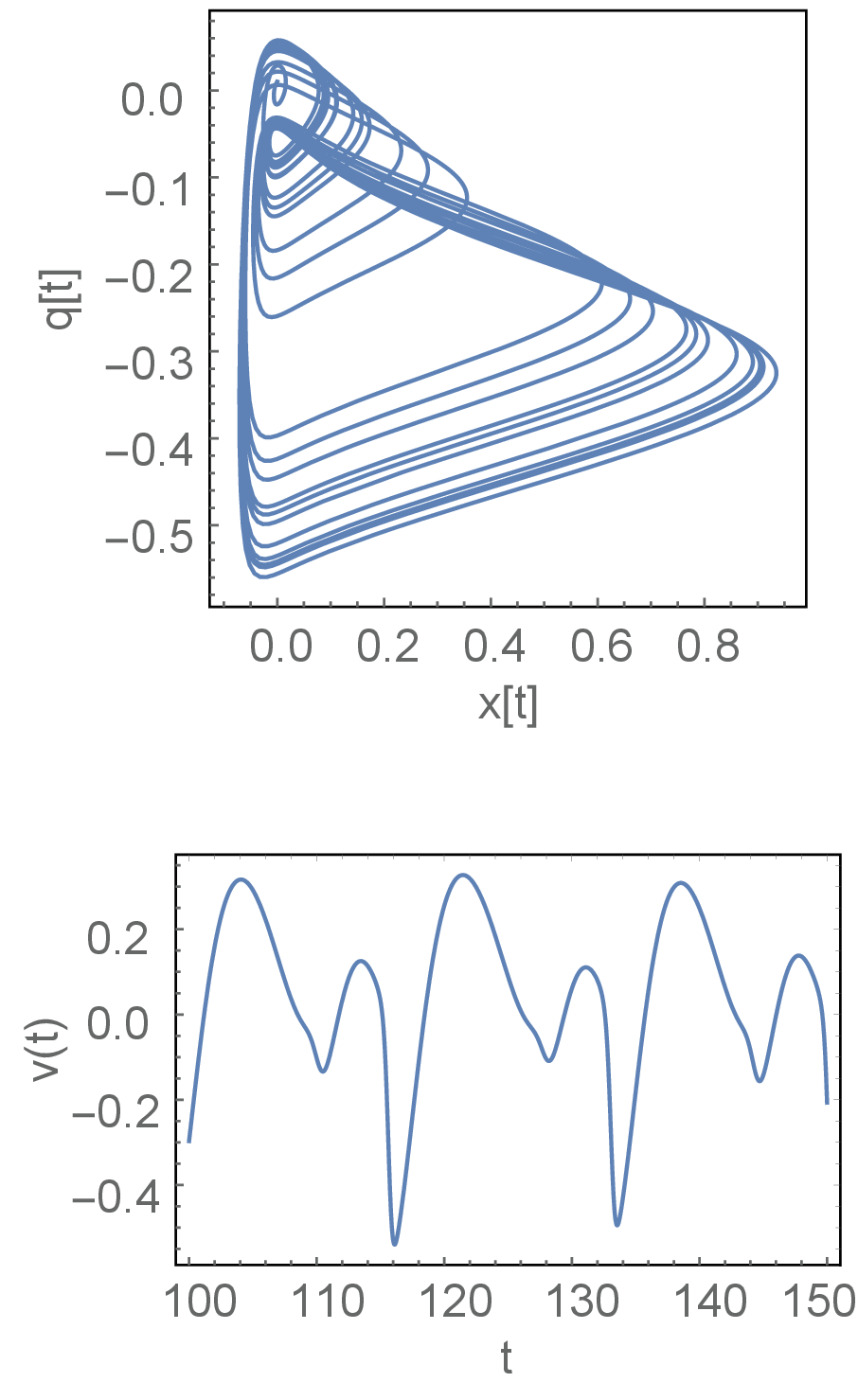}}
\caption{\small On the top, parametric plots of the deviation from equilibrium of the charge $q(t)$ vs.\ capacitor plate separation $x(t)$. The initial state is $x(0) = \dot{x}(0) = 0, q(0) = 0.01$. The bottom plots show the corresponding deviation from equilibrium of the voltage $v(t)$ as a function of the time. See the text (\Sec{sec:numerical}) for the values of the parameters corresponding to each case.\la{fig:trajectories}}
\end{figure} 

Figures \ref{fig:trajectories} and \ref{fig:pspectrum} refer to the following four choices of parameters:

\begin{enumerate}[label=(\alph*)]

	\item $ \gamma = 0.1$, $\Gamma_0 = 0.1$ ,  $\beta = 1$, and $\eta = -0.48$.  This corresponds to a low dissipation regime with a negative feedback parameter. The system rapidly tends to the fixed point.

	\item $ \gamma = 1$, $\Gamma_0 = 0.3 $,  $\beta = 10$, and $\eta = 0.86$.  This corresponds to a moderate dissipation regime with a subcritical feedback parameter. Sinusoidal and exponentially damped oscillations of $x , q$ and $v$ approach the fixed point.

	\item $ \gamma = 1$, $\Gamma_0 = 0.5 $,  $\beta = 10$, and $\eta = 1.14$.  This corresponds to a moderate dissipation regime with an over critical feedback parameter. The system slowly approaches a limit cycle with approximately sinusoidal voltage oscillations. The fundamental frequency and its harmonics are clearly visible in the power spectrum plot.  This corresponds to a weakly nonlinear regime, as described in \cite{SO}.

	\item $ \gamma = 0.5$, $\Gamma_0 = 0.1 $,  $\beta = 50$, and  $\eta = 4.52$.  This corresponds to a moderate dissipation regime, high $\beta$, and highly over-critical feedback parameter.  The attractor projected on the $(x,q)$ plane [see \Fig{fig:trajectories}(d)] appears to exhibit a self-similar (fractal) structure.  This and the noisy component of the corresponding power spectrum (see second plot in \Fig{fig:pspectrum}) are indicative of chaos.

\end{enumerate}

Chaotic behavior is to be expected in a nonlinear dynamical system with more than two degrees of freedom, but previous work on the electron shuttle (another electromechanical self-oscillator) has reported only limit-cycle behavior \cite{shuttle1, shuttle2}.  The chaotic regime of the LEC could be of some interest for biophysics, since the coexistence of regular and chaotic dynamics is a feature of biological systems at various levels of complexity; see, e.g., \cite{heart}.  The numerical results presented here are far from an exhaustive characterization of the mathematical properties of the LEC as a dynamical system.  Since the focus of the present work is on the physical interpretation and application of this new model, such an analysis is left for future work.

\begin{figure} [t]
\addtocounter{subfigure}{2}
\centering
	\subfigure[]{\includegraphics[height=0.25 \textwidth]{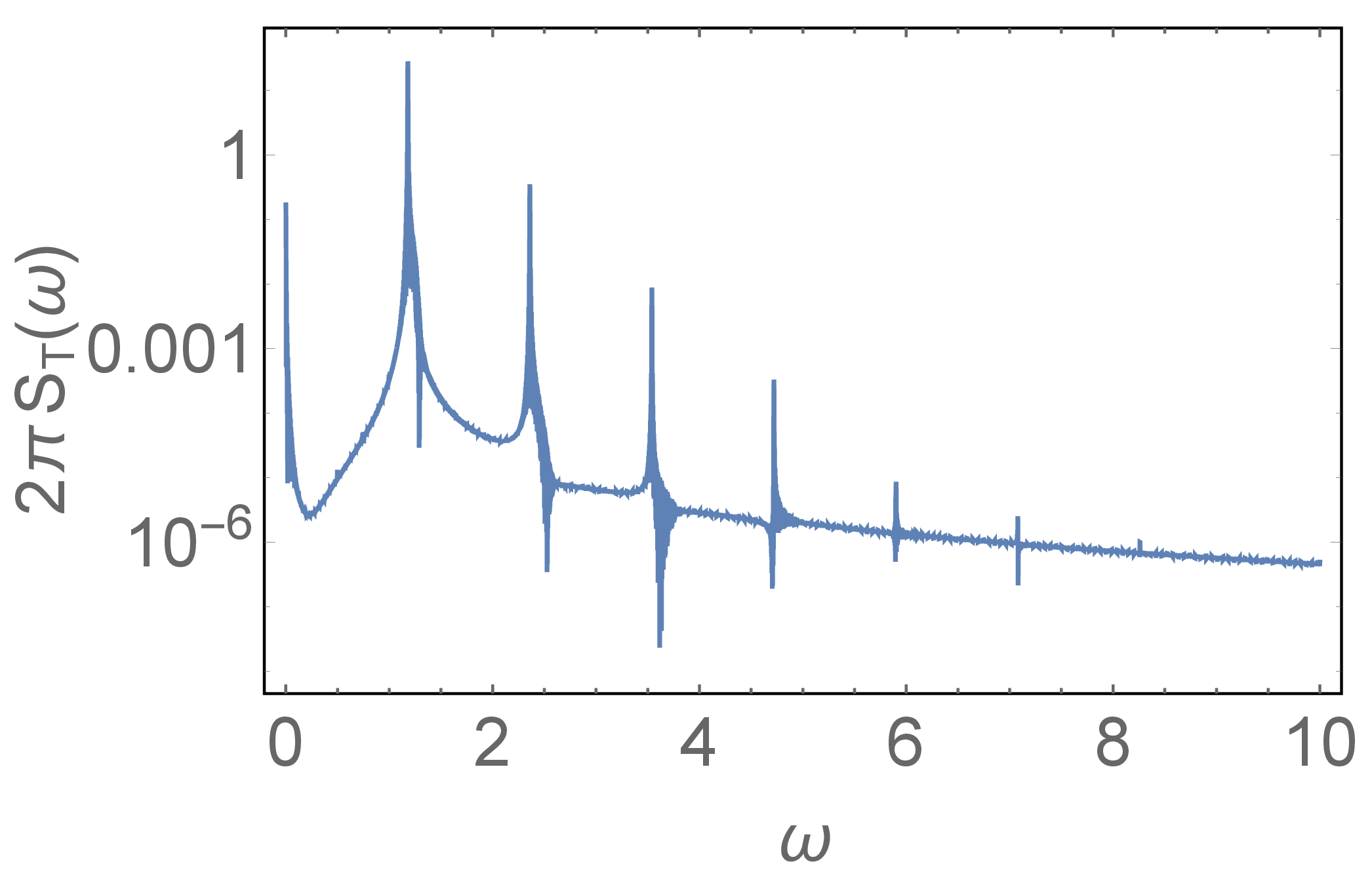}} \hskip 1.5 cm
	\subfigure[]{\includegraphics[height=0.25 \textwidth]{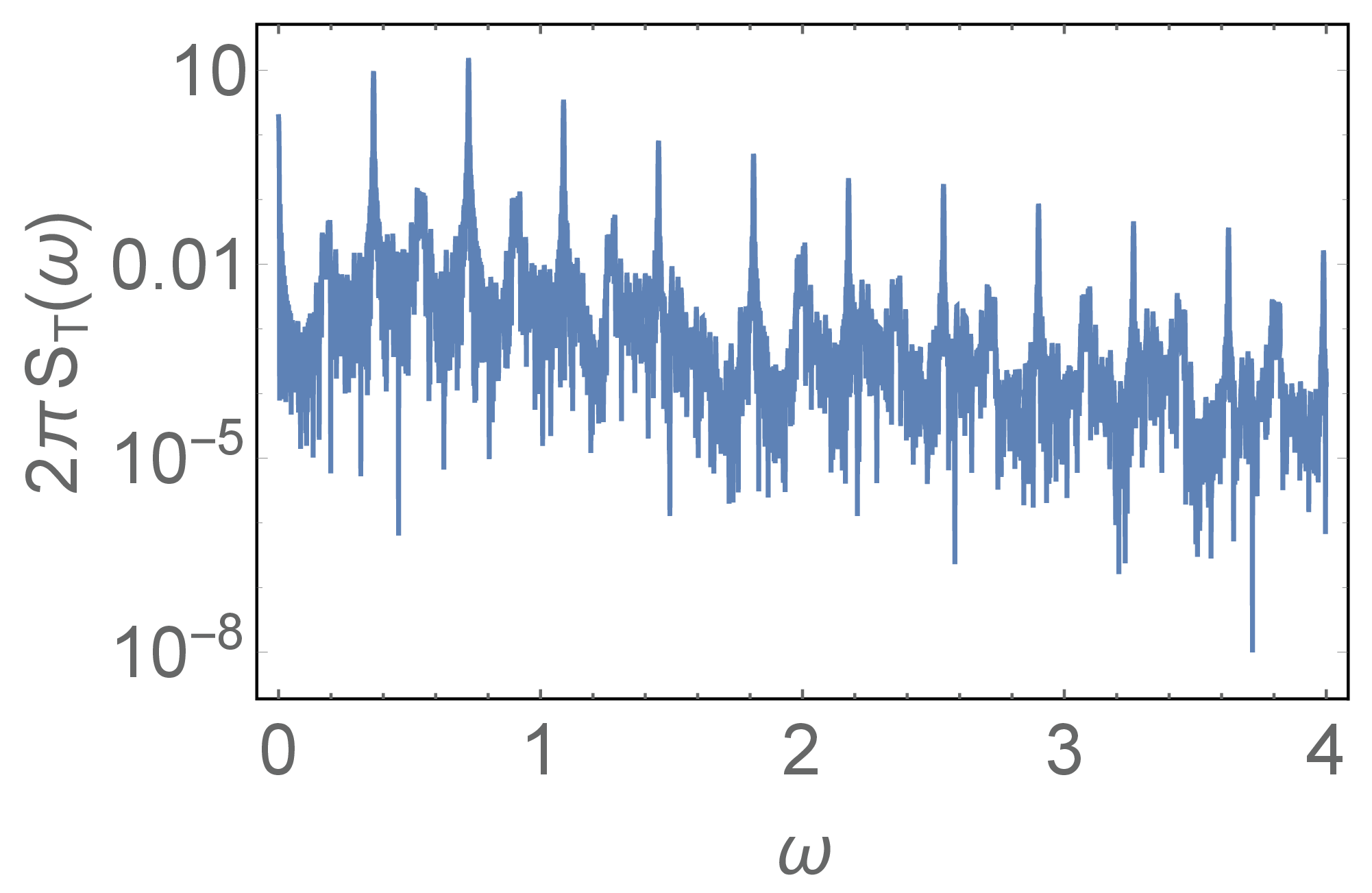}}
\caption{\small Power spectrum for the choices of parameters corresponding, respectively, to cases (c) and (d) in the text (\Sec{sec:numerical}).  These spectra are computed for a time interval $T= 1600$.\la{fig:pspectrum}}
\end{figure}


\section{Electromotive force}
\la{sec:emf}

We have demonstrated that positive feedback between the mechanical and the electrical degrees of freedom can destabilize the equilibrium configuration of the LEC, producing either regular self-oscillation or chaotic motion.  In either case, this is a manifestation of the extraction of work, represented by the integral of \Eq{eq:dW} over a complete cycle of the LEC's motion.  In this section we discuss how that work can be used to generate an emf, and therefore to pump an electrical current.

In classical electrodynamics, the emf is often equated with the circulation of the electric field,
\be
{\cal E} = \oint_{\cal C} \vv E \cdot d \vv s = - \frac 1 c \frac{d}{dt} \int_{\cal A} \vv B \cdot d \vv a ,
\la{eq:cemf}
\ee
where $\cal C$ is a closed path along the relevant circuit and $\cal A$ is the area that this path encloses.  But many devices, such as batteries, photovoltaic cells, fuel cells, and thermoelectric generators, can actively drive current along a closed circuit $\cal C$ even though the circulation in \Eq{eq:cemf} is negligible because there is no coherent, time-varying net magnetic flux through $\cal A$.

If the electric field is irrotational (i.e., $\Del \times \vv E \equiv 0$), there exists a scalar potential $\phi$ such that
\be
\vv E (t, \vv r) = - \Del \phi (t, \vv r) .
\ee
The power that the electric field exerts on the current contained within a volume $\cal V$ can be expressed in terms of the current density $\vv J (t, \vv r)$ as
\be
P = \int_{\cal V} \vv E \cdot \vv J \, d^3 r = - \int_{\cal V} ( \Del \phi ) \cdot \vv J \, d^3 r = - \oint_{\cal S} \phi \vv J \cdot d \vv a + \int_{\cal V} \phi (\boldsymbol \nabla \cdot \vv J ) \, d^3 r ,
\la{eq:parts}
\ee
where $\cal S$ is the two-dimensional, closed boundary of the volume $\cal V$.  If $\cal V$ contains all points at which $\vv J \neq 0$, the surface term in \Eq{eq:parts} vanishes.  Charge conservation implies the continuity condition
\be
\Del \cdot \vv J = - \frac{\partial \rho}{\partial t} ,
\la{eq:cont}
\ee
where $\rho(t,\vv r)$ is the local charge density.  Combining Eqs.\ \eqref{eq:parts} and \eqref{eq:cont}, we obtain
\be
P =-  \int_{\cal V} \phi \frac{\partial \rho}{\partial t} \, d^3 r .
\la{eq:power}
\ee

In a discharging capacitor, we have that $\partial \rho / \partial t < 0$ in the conducting plate with $\phi = \phi_+$, and the opposite $\partial \rho / \partial t > 0$ in the conducting plate with $\phi = \phi_-$.  Let $\cal V = \cal V_+ + \cal V_-$, where $\cal V_\pm$ are the volumes of the respective plates.  Then \Eq{eq:power} implies that
\be
P = - \phi_+ \int_{\cal V_+} \frac{\partial \rho}{\partial t} \, d^3 r  + \phi_- \int_{\cal V_-} \frac{\partial \rho}{\partial t} \, d^3 r = VI, 
\ee  
where $V = \phi_+ - \phi_-$ and the current $I$ is the absolute value of the integral of $\partial \rho / \partial t$ over either $\cal V_+$ or $\cal V_-$.  In the steady state, this positive power delivered by the electric field to the moving charges is matched by the negative power $I R^2$ dissipated by the internal friction of the load resistance $R$.  Note that, in this case, the charge does not move along a closed path and no emf is involved.  In the scheme described in the Appendix, this corresponds to a {\it passive} electrical system.

For a stationary charge density $\partial \rho / \partial t \equiv 0$ and \Eq{eq:power} implies that $P \equiv 0$.  However, if $\partial \rho / \partial t \neq 0$ then current may be driven along a closed path, even though the circulation of the electric field in \Eq{eq:cemf} vanishes.  This is the case, e.g., in cyclotrons and other particle accelerators, where a charged bunch goes around in such a way that its motion is synchronized with the switching of a localized electric potential \cite{cyclotron}.

Note that the short-range, repulsive interactions between particles (a quantum effect not properly described in terms of the classical $\vv E$ field) can be incorporated into this analysis by including a chemical potential term $\mu_n$ for each species $n$ with particle charge $q_n$.  In terms of the electrochemical potential
\be
\Phi_n = q_n \phi + \mu_n ,
\ee
\Eq{eq:power} can be generalized to
\be
P_n =-  \int_{\cal V} \Phi_n \frac{\partial \rho_n}{\partial t} \, d^3 r ,
\la{eq:pwr-n}
\ee
where $P_n$ is the power transmitted to the current of particle $n$, with number density $\rho_n$.  Note that $P_n > 0$ is incompatible with stationary densities ($\partial \rho_n / \partial t \equiv 0$).

Now let us consider the case of the self-oscillating LEC.  Suppose that the negatively charged plate is fixed and that the positively charged plate oscillates in $X$.  In the contraction phase ($\dot X < 0$), the positive charge moves out of a region of higher potential $\phi$ (corresponding to the separation $X$ at time $t$) and into a region of lower potential $\phi - d \phi$ (corresponding to the separation $X - dX$ at time $t + dt$).  In other words, in the vicinity of the positive plate the charge density is leaving a region of higher potential and entering a region of lower potential, which implies that $P > 0$ in \Eq{eq:power}.

On the other hand, in the expansion phase ($\dot X > 0$), the positive charge is at each moment moving out from $X$ to $X + dX$, but the region into which the plate is moving is electrically screened.  For the ideal double-plate capacitor, this would give $P = 0$ in \Eq{eq:power}.  In a more realistic setting, we expect $P < 0$, but the power lost in the expansion phase will be less than the power gained in the contraction phase.  If we define $\bar P$ as the average of the power over a full period $\tau$ of the LEC's self-oscillation,
\be
\bar P = \frac 1 \tau \int_0^\tau P(t) dt ,
\ee
we therefore can obtain $\bar P > 0$, because the symmetry between contraction and expansion is broken by the fact that the electric field is large inside the capacitor and small outside of it.

This means that the mechanical oscillation of an isolated elastic capacitor must die out quickly, since the mechanical oscillation leads to a net acceleration of the charges, and this energy must be dissipated within the capacitor.  On the other hand, the LEC connected to an external circuit can generate an effective emf,
\be
{\cal E} = \frac{\bar PT}{\bar Q} > 0 ,
\la{eq:emf}
\ee
where $\bar Q$ is the total charge driven around the closed circuit during a macroscopic time $T \gg \tau$.  That is, the self-oscillating LEC is an electrically {\it active} device that can pump an electric current through an external load connected to its terminals (see the Appendix).  We therefore expect the LEC model to be applicable to the description of the operation of various active electrical devices in which the emf cannot be described in terms of \Eq{eq:cemf}.  Note that the quantity $\bar P T$ in the numerator of \Eq{eq:emf} is equal to the mechanical work $W$ done during time $T$ by the self-oscillating LEC's nonconservative force [see \Eq{eq:dW}], minus some loss due to internal frictions that make the efficiency of conversion from mechanical to electrical work less than 1.

In a particle accelerator, the motion of the charges and the modulation of the electrical potential must be externally synchronized to achieve $\bar P > 0$.  But in the LEC both of them are controlled by the same self-oscillation in a way that generates an emf autonomously.  Note that the emf of \Eq{eq:emf} will, in general, depend on the average current $I = \bar Q / T$ drawn from the active device.  The battery's emf is usually measured under open-circuit conditions, which corresponds to taking $I \to 0$ (in which case also $\bar P \to 0$).  The decrease in $\cal E$ as $I$ increases is then treated phenomenologically as an ``internal resistance''.  This is similar to the behavior of hydraulic pumps, in which the ``head rise'' (analogous to the emf) begins to decrease as the flow rate (mass of water drawn per unit of time) increases; see, e.g., \cite{pumps}.

The problem of understanding the dynamics of the generation of an emf in the absence of a macroscopic and time-varying magnetic flux applies to a variety of microscopic energy transducers, including batteries, fuel cells, solar cells, and thermoelectric generators.  The proposal that we have worked out here of the LEC as a self-oscillating charge pump is qualitatively similar to what we had previously proposed to the solar cell \cite{AGJ}. There, however, we failed to recognize the fundamental difference between the operation of a capacitor and that of a battery, considered as power sources.  This led us to a mistaken contrast between the solar cell as an active system and the battery as a passive one.  On the application of the LEC model to the battery, see \cite{battery}.

In plasma physics, the role of EDLs in the acceleration of particles has aroused controversy, both because the formation of a double layer is difficult to describe using the equations of magnetohydrodynamics and because the electric field of a static EDL is conservative and therefore cannot act as a particle accelerator \cite{Bryant}.  On the other hand, experiments show that plasma double layers oscillate with both slow and fast frequency components, with the fast component being close to the plasma frequency \cite{Torven, Volwerk}.  We therefore expect that our LEC model may be directly relevant to the dynamics of plasma double layers.

Experiments in sonoelectrochemistry have found that acoustic driving can pump current through an EDL \cite{Compton}.  In a recent review of the subject, this effect has been described as ``a great enhancement of mass transfer of electroactive species from the bulk solution to the electrode surface through the double layer'', induced by an ultrasound signal \cite{sonoelectrochem}.  The self-oscillation of an EDL can be used directly to pump a current, as proposed in \cite{battery} in the context of batteries.  The case of ion transport in biological cells is more involved, since the membrane's ion channels have complex molecular structures that make them chemically selective.  Nonetheless, since the results obtained here establish that the self-oscillation of an EDL can actively pump an electric current in a way that is thermodynamically irreversible and compatible with classical electrodynamics, we expect that this can help to clarify some of the basic obscurities in our present understanding of the role of the electric field in the transport of charges in biological systems.  On that question see, e.g., \cite{Eisenberg} and references thererin.


\section{Self-sustained traveling waves}
\la{sec:traveling}

Partenskii and Jordan have shown the importance of incorporating lateral flexibility into electromechanical models of EDLs \cite{Partenskii1996, Partenskii2001, Partenskii2005, Partenskii2009, Partenskii2011}.  Correspondingly, various forms of active matter may be modeled by an extension of the LEC in which the capacitor plates are not perfectly rigid.  An important example of this is the biological cell membrane, whose mechanical elasticity can be described by various moduli \cite{moduli}.  Such membranes can therefore support traveling waves, with possibly complicated dispersion laws.

A patch of membrane can be treated as a capacitor charged by biophysical processes involving active channels (pumps), and discharged by passive channels.  Both types of channels are controlled by electromechanical forces \cite{AK, membrane-forces}, yielding the feedback that can generate self-oscillations in the LEC model.  Locally excited oscillations can propagate along the plane of the double layer as self-sustained waves, fed by energy from the underlying chemical process.  This energy is necessary to compensate the wave damping, which must be considerable in a soft-matter system.  We consider it plausible that such self-sustained waves may transport information and electrical energy in biological systems.

As an illustration of how this mechanism could work, consider a spatially uniform model admitting only transverse waves and treated within the linear approximation.  The double-layer plane is parametrized by spatial $(x, y)$ variables.  The field $\psi(x,y; t)$ represents the local variation of the inter-layer distance, previously denoted by $X(t) - X_0$. Another field, $n(x,y;t)$, represents the variation of the local density of the relevant charge carriers, replacing $Q(t)- Q_0$ in the LEC model in \Sec{sec:model}.  This gives us a wave equation with dissipation for $\psi$, coupled to a reaction-diffusion equation for $n$ (on reaction-diffusion systems and their applications to pattern formation in active media, see \cite{Cross} and references therein).  In the linear approximation
\bea
\left[ \frac{\partial^2}{\partial t^2} - {c_s^2} \left( \frac{\partial^2}{\partial x^2} + \frac{\partial^2}{\partial y^2} \right) + \Omega_0^2 \right] \psi(x,y;t) &=& -\gamma \cdot \frac{\partial}{\partial t}\psi(x,y;t) - a \cdot n(x,y;t) , \la{eq:wave} \\
\frac{\partial}{\partial t}n(x,y;t) - D \left( \frac{\partial^2}{\partial x^2}+\frac{\partial^2}{\partial y^2} \right) n(x,y;t) &=& -\Gamma_0 \cdot n(x,y;t) + b \cdot \psi(x,y;t) . \la{eq:reaction}
\eea
Here $c_s$ is the phase velocity of the free linear waves and $D$ is the diffusion constant for charge carriers.  The parameters $\Omega_0 , \gamma, \Gamma_0, a, b$ have a similar meaning to the corresponding symbols used in \Sec{sec:model} [see \Eq{eq:params}].

We are interested in solutions to this system of coupled equations corresponding to self-sustained plane waves:
\be
\psi(x,y;t) = \Psi \cos[ \omega(k) t - k_x x - k_y y ], \quad n(x,y;t) = N \sin[ \omega(k)  t - k_x x - k_y y + \alpha ], \quad  k = \sqrt{k_x^2 + k_y^2} .
\la{eq:plane}
\ee
Inserting \Eq{eq:plane} into Eqs.\ (\ref{eq:wave}) and (\ref{eq:reaction}) one obtains the following relations for the parameters of this solution.

\begin{enumerate}[label=\arabic*.]

	\item Dispersion law: \be \omega(k) =  \sqrt{\bar c ^2 k^2  + \bar \Omega ^2}, \quad \bar c = \sqrt{c_s^2 + D \gamma}, \quad \bar \Omega = \sqrt{\Omega_0^2 + \gamma\Gamma_0} .  \la{eq:dispersion} \ee

	\item Phase-shift parameter $\alpha$: \be \tan \alpha = \frac{D k^2 + \Gamma_0}{\omega(k)} . \la{eq:alpha} \ee

	\item $N/\Psi$ ratio: \be \frac N \Psi = \frac{\gamma \omega(k)}{a \cos \alpha} . \la{eq:ratio} \ee
	
\end{enumerate}

The last equation, of the form
\be
\omega(k)^2 + \left( D k^2 + \Gamma_0 \right)^2 = \frac{ab}{\gamma} ,
\label{eq:kvector}
\ee
gives the unique value  $k_m$ of the magnitude of the wave vector for self-sustained plane waves. Note that $k_m$ exists only for strong enough coupling between the electro-mechanical and the chemical components of the system, characterized by the product $ab$. The threshold for its existence is given by the inequality
\be
ab > \gamma \cdot \left( \bar \Omega^2 +  \Gamma_0^2 \right) .
\la{eq:threshold}
\ee
The critical value $k_m$ corresponds to the bifurcation point in the LEC model in \Sec{sec:model}, given by the equality $\eta = 1$ in \Eq{eq:eta}. It means that the waves with lower $k$ are exponentially amplified while those with higher $k$ are exponentially damped. Due to nonlinear effects, the waves with $0 \leq k \leq k_m$ are stabilized and can form self-sustained, propagating wave packets.

The dispersion law of \Eq{eq:dispersion} implies that one can generate localized wave packets moving at arbitrarily low group velocities.  It is amusing to note that the resulting mathematical description is analogous to that of the relativistic de Broglie matter waves in quantum mechanics (see, e.g., \cite{Davydov}), with $\bar c$ playing the role of the speed of light and $\bar \Omega$ that of the mass (in units in which Planck's constant is set to $\hbar = 1$).  The fast oscillation of the separation between the flexible plates, with frequency $\bar \Omega$, is then analogous to the unobservable {\it Zitterbewegung} (``trembling motion'') of the relativistic matter waves \cite{Zitter}.

In \Sec{sec:SO} we discussed how the LEC could self-oscillate in a weakly nonlinear regime, with approximately sinusoidal motion, or in a strongly nonlinear (relaxation oscillation) regime, with a frequency controlled by the damping rates.  Those respective regimes have their analogs in the model of self-sustained traveling waves given by Eqs.\ \eqref{eq:wave} and \eqref{eq:reaction}.  There is a regime in which waves travel almost adiabatically, carrying substantial electromechanical energy with them.  On the other hand, there is a strongly over damped regime in which $\bar c$ and/or $\bar \Omega$ in \Eq{eq:dispersion} are controlled by the damping rates, rather than by the conservative parameters $c_s$ and $\Omega_0$.  In the latter case, an incoming pulse does not impart significant electromechanical energy, since most of that is rapidly dissipated.  Instead, the traveling wave is sustained because the incoming pulse absorbs energy from the local electric discharging.\footnote{This can be pictured as analogous to the ``Mexican wave'' behavior of human crowds in sports stadiums \cite{Mexican-wave}.  No significant energy is transmitted among individuals as the wave moves through the crowd.  Rather, the wave is sustained entirely by the internal energy of the participating individuals.}  As before, we note that this regime may be particularly important in soft-matter systems.

Let us now show that, if this model were applied to neural axons, the power supplied by the charged membrane would suffice to produce a self-sustained wave distinguishable from thermal noise.  The driving power density per unit surface $P_d$ should be of the same order of magnitude as the electric power density corresponding to the typical current through the membrane (which is given by active ion pumping and driven by metabolic processes).  That power is $V^2_0 /R$, where $V_0 \simeq 0.1$ V is the typical membrane voltage and $R \simeq 10^{-3} \Omega \, \hbox{m}^2$ is the minimal specific resistance of the membranes \cite{Izhikevich}. Thus,
\be
P_d \simeq \frac{V_0^2}{R} \simeq 10 ~ \hbox{W} / \hbox{m}^2 .
\la{eq:Pd}
\ee 
We take the axon of the neuron to be a hollow cylinder whose surface is a bimolecular lipid membrane.  The diameter of that cylinder is $d \gtrsim 1 ~\mu$m \cite{Siegel-axon}.  The duration of the spike at a fixed point along the axon is $\Delta t \simeq 1$ ms \cite{Izhikevich}.  The nerve conduction velocity is $v \simeq 50$ m/s \cite{Siegel-v}.  The effective width of the spike is therefore
\be
\ell = v \cdot \Delta t \simeq 5 ~\hbox{cm} .
\la{eq:ell}
\ee
The effective surface area of the spike is then
\be
S = \pi d \ell \gtrsim1.5 \times 10^{-7} ~\hbox{m}^2 .
\la{eq:S}
\ee
Combining Eqs.\ \eqref{eq:Pd} and \eqref{eq:S}, we estimate the driving power supplied to the excited surface to be
\be
P_S = P_d \cdot S \gtrsim 10 ~ \hbox{W} / \hbox{m}^2 \times 1.5 \times 10^{-7} ~\hbox{m}^2 = 1.5 ~ \mu \hbox{W} .
\ee

Let $P_S^{\rm th}$ be the power per unit surface supplied by the thermal environment to the few (order 1) mechanical degrees of freedom that support the spike.  The condition for being able to distinguish the spike signal from thermal noise is
\be
P_S \gg P_S^{\rm th} \simeq \frac{k_B T}{\tau} ,
\la{eq:noise}
\ee
where $k_B$ is the Boltzmann constant, $T$ the ambient temperature, and $\tau$ the thermal relaxation time for the bimolecular lipid membrane.  Thus the plausibility of our model for neural signaling requires that
\be
\tau \gg \frac{k_B T}{P_S} \lesssim \frac{10^{-20} ~\hbox{J}}{1.5 ~ \mu \hbox{W}} \simeq 10^{-14} ~\hbox{s} .
\la{eq:scales}
\ee
The typical time scale for heat transfer along the bimolecular lipid membrane, as reported in \cite{Smits}, is $\tau \simeq 1$ ps, fulfilling the condition of \Eq{eq:scales}.


\section{Discussion}
\la{sec:discussion}

We have proposed a simple dynamical system, the LEC, whose operation depends only on electrostatic and mechanical forces, plus ohmic conductance.  Feedback between the mechanical separation $X$ and the capacitor charge $Q$ can cause this system to self-oscillate, making it into an AC generator.  nonlinearity is needed only in the mechanical response of the elastic capacitor for large displacements away from the equilibrium separation, which is obviously a physically realistic property.  In \Sec{sec:numerical} we have also used a nonlinear relation between $X$ and the capacitor's internal resistance $R_i$ [see \Eq{eq:res1}], but this is not necessary to obtain the active dynamics of the LEC.

The simplicity of the LEC may be useful in the design of cheap electromechanical clocks and other oscillators.  It may also provide a more elementary and realistic description of the dynamics accounting for the extraction of work that characterizes active matter, as stressed in \Sec{sec:intro}.  For instance, the study of excitable membranes in biophysics has, so far, focused on mathematical descriptions of their function rather than on a realistic understanding of their physical mechanism.  Thus, the FitzHugh-Nagumo model \cite{FitzHugh, Nagumo} uses an inductor and a tunnel diode, electronic elements not found in living matter.  The simple Lapicque integrate-and-fire model \cite{Abbott}, like the more sophisticated Hodgkin-Huxley model \cite{Hodgkin-Huxley}, invokes a current source without attempting to describe the active dynamics that generates and sustains that current.

The LEC model may point towards a more realistic physical description of excitable membranes, helping to realize the project, which Babakov, Ermishkin, and Lieberman suggested already in 1966, of accounting for the ``activity of the cell membrane'' in terms of its electromechanical properties \cite{Babakov}.  The importance of the mechanical properties of such membranes is now well established \cite{AK, membrane-forces}.  The LEC model shows how the underlying electrochemical disequilibrium across the membrane can power a sustained (i.e., engine-like) generation of mechanical work.  That work can then be used to generate the emf that pumps the active currents seen in living cells.  We have also sketched in \Sec{sec:traveling} how an extension of the LEC model, which includes a mechanical flexibility of the capacitor plates, may provide a realistic description of neural signaling and electrical energy transport based on self-sustained traveling waves.

Our research has been motivated in large part by the modern ``quantum thermodynamics'' approach to understanding the dynamics in time by which an open system irreversibly extracts work from an underlying disequilibrium (see, e.g., \cite{QT} and references therein), even though everything in the present treatment of the LEC has been classical.  For recent work in which the microphysics of the generation an emf by a triboelectric generator is considered in terms of an open quantum system, see \cite{tribo}.

Note that the active LEC is a ``dissipative structure'' in the sense in which Prigogine and his collaborators introduced that term into the theory of nonequilibrium thermodynamics \cite{Prigogine}.  That approach, like Haken's ``synergetic'' treatment of similar questions \cite{Haken}, was met with strong objections (see, e.g., \cite{Anderson}), which in our view were justified by its inability to describe the dynamics in time of such structures operating as {\it cyclical engines}.  Prigogine, Haken, and other theoreticians had inherited that conceptual limitation from Onsager's formulation of nonequilibrium thermodynamics (including active systems like thermoelectric generators) in terms only of thermodynamic potentials and their gradients \cite{Onsager}.

What is qualitatively novel in our description, compared to that Onsagerian approach, is our focus on the feedback dynamics giving raise to the active nonconservative force of \Eq{eq:force}.  It is this force that is responsible for the work output of \Eq{eq:dW}, which in turn is the physical explanation of the dynamics in time illustrated by the numerical results of \Sec{sec:numerical}.  On why this approach is needed to describe active matter in a physically realistic manner, see also the discussion in the Appendix.

With some natural generalizations, the LEC's equations of motion define a class of dynamical systems with a broad range of possible applications.   The presence of EDLs in many active systems of interest in solid-state, plasma, chemical, and biological systems suggests that the LEC could find a range of interesting applications, some of which we mentioned in \Sec{sec:intro}.   We therefore hope that the simple model that we have presented here may open new avenues towards a better understanding of active systems across various areas of pure and applied science.

One question that evidently calls for careful investigation is the microphysical derivation of a realistic potential $U_m(X)$ in \Eq{eq:U} for the EDL in each of the active systems of interest.  Another is the inclusion of thermal noise in a consistent statistical treatment of LEC dynamics that distinguishes between dissipation by viscous damping of the energy in the oscillation of $X$ and the extraction from it of work by pumping of zero-entropy currents (on this problem see Sec.\ III D in \cite{battery}, as well as the treatment of the single-electron shuttle in \cite{Strasberg}).  We have also left for future research the details of the dynamics of nonrigid LECs and the simulation of the corresponding traveling waves proposed in \Sec{sec:traveling}.  Much analytical and numerical work remains to be done in order to fully understand the possible dynamics that can be obtained from the LEC model and its extensions, but we are confident that the results presented here already offer significant lessons for the better understanding of active matter.


\begin{acknowledgments} We thank Elizabeth von Hauff for extensive discussions about the model presented in this article and its possible applications. A.J.\ also thanks Esteban Avenda\~no, Diego Gonz\'alez, Mavis Montero, and Roberto Urcuyo for discussions of the dynamics of electrochemical double layers.  R.A.\ was supported by the International Research Agendas Programme (IRAP) of the Foundation for Polish Science (FNP), with structural funds from the European Union (EU). D.G.-K.\ was supported by the Gordon and Betty Moore Foundation as a Physics of Living Systems Fellow (Grant No.\ GBMF45130).  AJ was supported by the Polish National Agency for Academic Exchange (NAWA)'s Ulam Programme (Project No.\ PPN/ULM/2019/1/00284). \end{acknowledgments}


\newpage


\appendix*
\counterwithin{figure}{section}
\renewcommand{\thefigure}{A\arabic{figure}}

\section{Passive vs.\ active systems}
\la{sec:active}

Electronics textbooks distinguish between {\it passive} components such as resistors, capacitors, inductors, and transformers on the one hand, and {\it active} components, such as transistors and operational amplifiers on the other.  The usual definition given is that an active component, unlike a passive one, can amplify the power that it receives from the circuit.  Evidently, this requires that the active component have an external source of power which is {\it not} the circuit on which it operates.  Horowitz and Hill add that active devices ``are distinguishable by their ability to make oscillators, by feeding from output signal back into the input'' \cite{HH}, i.e., to self-oscillate.  According to this classification, the ordinary capacitor is passive while the LEC is active.

Compare this to the definition of ``active matter'' given in statistical mechanics and in soft condensed-matter physics, as matter composed of elements that can consume and dissipate energy, in the process executing systematic movement \cite{Ramaswamy}.  The recent literature on active matter usually accounts for the energy consumed by those elements from an underlying thermodynamic disequilibrium, but it does not treat the dynamics of work extraction in a physically realistic way.  As we have argued in \Sec{sec:intro}, this is because it does not consider the feedback dynamics that produces the nonconservative force that drives the piston or turbine through which work is extracted.

\begin{figure}[t]
\centering
	\subfigure[]{\includegraphics[height=0.24 \textwidth]{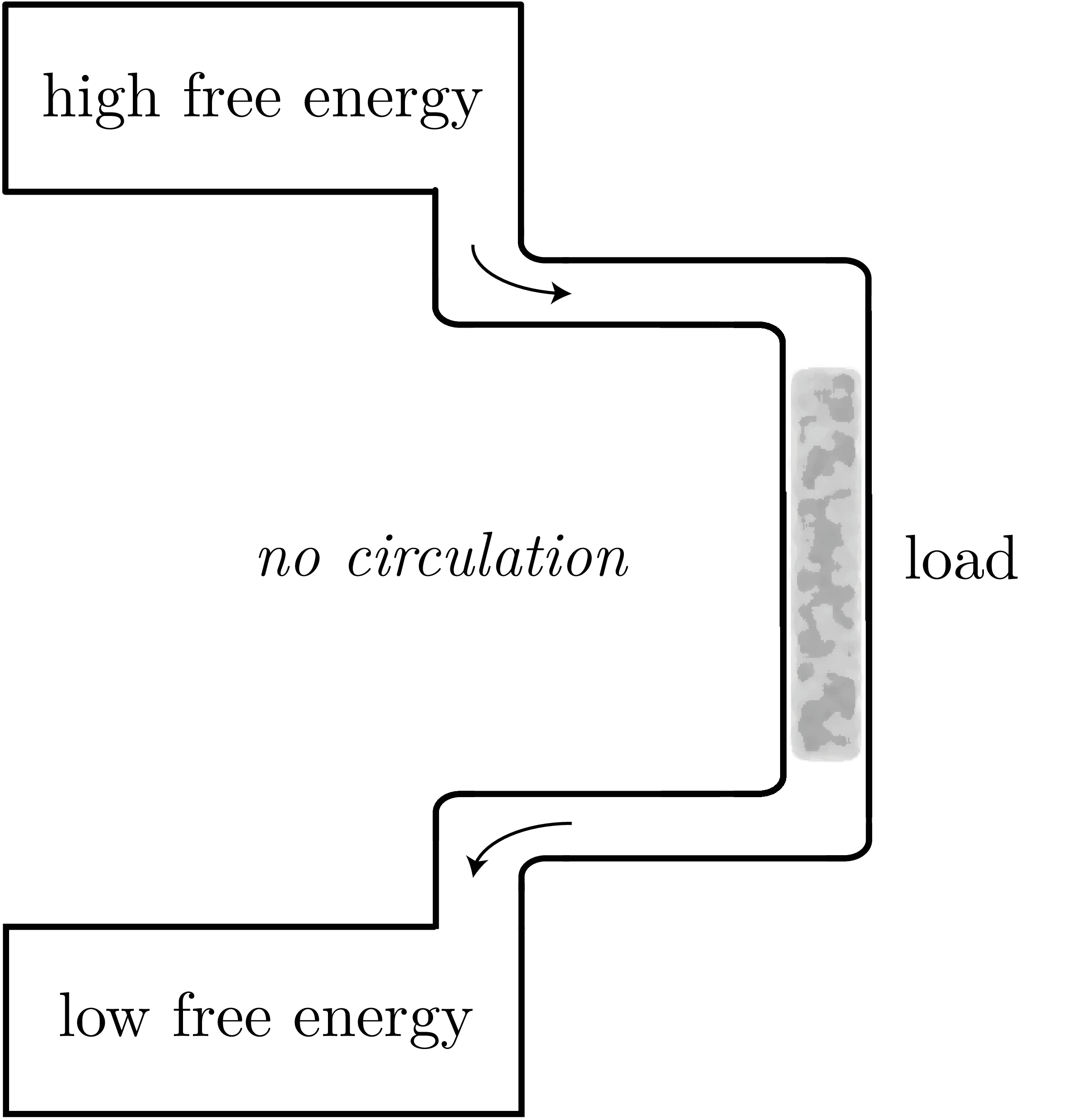}} \hskip 2 cm
	\subfigure[]{\includegraphics[height=0.24 \textwidth]{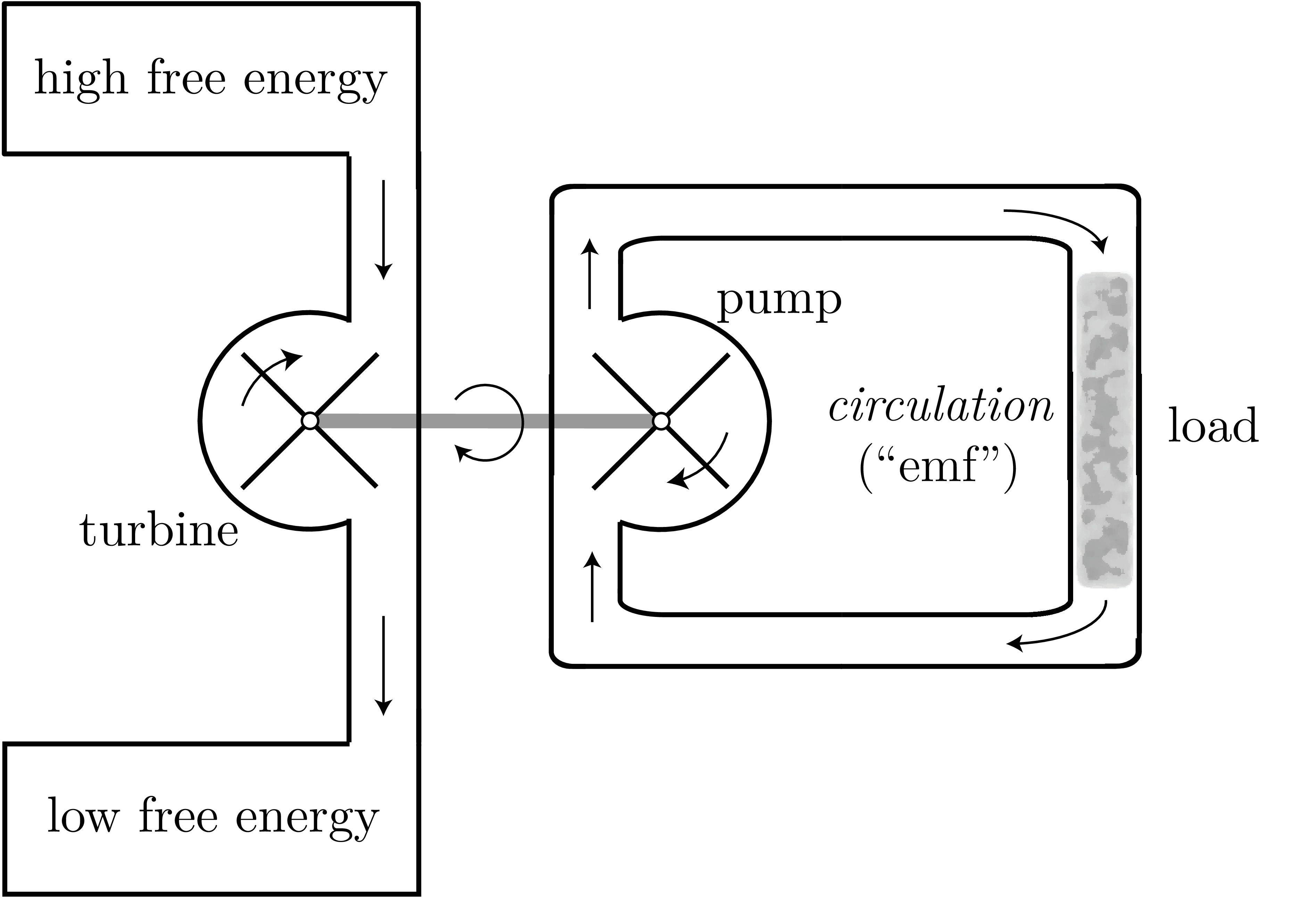}}
\caption{(a) General conceptual scheme for a passive device that consumes free energy from an external source.  This energy is dissipated in the load and the system cannot maintain any circulation of the flow.  (b) General conceptual scheme for an active device that uses the free energy consumed from an external source to perform sustained work, represented here as the driving of a pump that maintains the circulation of the flow in the circuit on the right.  For an electrical device, the integral per unit charge of the active, nonconservative force that drives that circulation is the emf.\label{fig:passive-active}}
\end{figure}

What we mean by the precise distinction between a passive system and an active one, as we have applied it throughout this article, is schematically illustrated in \Fig{fig:passive-active}.  A passive system, represented by \Fig{fig:passive-active}(a), can consume and dissipate free energy, but it cannot use this to perform sustained work or to generate an emf.  An active system, represented in \Fig{fig:passive-active}(b), uses some of the free energy that it consumes to generate an active, nonconservative force, which can be used to pump a flow against an external potential or to sustain a circulation.  For an electrical device, the integral of the force per unit charge that produces that circulation corresponds to the emf.

Note that \Fig{fig:passive-active} shows the work being extracted by a rotor or turbine, rather than by a self-oscillating piston like the plate separation $X$ of the LEC.  The mathematical description of the rotor's motion appears different to that of the piston; one may describe it as a ``self-rotation'' (see, e.g., \cite{autorotation}).  However, from a thermodynamic standpoint self-rotors and self-oscillators are very similar.  In both cases work is extracted irreversibly by the open system coupled to an external disequilibrium.  There is a positive feedback between the macroscopic motion of the rotor or oscillator and the state of the working medium, giving rise to the active nonconservative force that accounts for the system's persistent generation of work.  In the case of self-rotation, this may be characterized as a circulatory force.  On the generation of work by rotors considered in the context of quantum thermodynamics, see \cite{rotors}.

Our model of the LEC fits into the scheme for an active system of \Fig{fig:passive-active}(b).  The voltage $V_0$ is an external disequilibrium that can be used to extract work via the mechanical self-oscillation of $X$ (see \Fig{fig:LEC}), which acts as piston.  In \Sec{sec:emf} we showed how that this self-oscillation can, in turn, generate an emf, pumping electrical current through the double layer.  This emf can do something that the voltage $V_0$ on its own cannot: drive current along a closed circuit when a load is connected to the LEC's plates.

It is our contention that mathematical treatments of off-equilibrium processes based entirely on thermodynamic potentials and their gradients can correctly describe passive systems, but not active ones.  The ``blind spot'' of theoretical physics surrounding the dynamics-in-time of engines and active systems in general has caused a great deal of conceptual confusion.  On this, see also the discussion in \Sec{sec:discussion}.


\begin{thebibliography}{99}


\bibitem{AL}
	A.~Lasia, 
	``Electrochemical Impedance Spectroscopy and its Applications'',
	in {\it Modern Aspects of Electrochemistry} {\bf 32}, eds.\ B.~E.~Conway, J.~Bockris and R.~White,
	(Boston: Springer, 2002), \href{https://doi.org/10.1007/0-306-46916-2_2}{pp.\ 143--248}

\bibitem{Hamann}
	C.~H.~Hamann, A.~Hamnett, and W.~Vielstich,
	{\it Electrochemistry}, 2nd ed.,
	(Weinheim: Wiley-VCH, 2007), sec.\ 5.2.2
	
\bibitem{EvH}
	E.~von Hauff,
	``Impedance Spectroscopy for Emerging Photovoltaics'',
	J.\ Phys.\ Chem.\ C \href{https://doi.org/10.1021/acs.jpcc.9b00892}{{\bf 123}, 11329} (2019)
	
\bibitem{McL}
	S.~McLaughlin,
	``The Electrostatic Properties of Membranes'',
	Annu.\ Rev.\ Biophys.\ Biophys.\ Chem.\ \href{https://doi.org/10.1146/annurev.bb.18.060189.000553}{{\bf 18}, 113} (1989)
	
\bibitem{Raadu}
	M.~A.~Raadu,
	``The physics of double layers and their role in astrophysics'',
	Phys.\ Rep.\ \href{https://doi.org/10.1016/0370-1573(89)90109-9}{{\bf 178}, 25} (1989)
	
\bibitem{Babakov}
	A.~V.~Babakov, L.~N.~Ermishkin, and E.~A.~Liberman,
	``Influence of Electric Field on the Capacity of Phospholipid Membranes'',
	Nature \href{https://doi.org/10.1038/210953b0}{{\bf 210}, 953} (1966)
	
\bibitem{Lauger}
	P.~L\"auger, W.~Lesslauer, E.~Marti, and J.~Richter,
	``Electrical properties of bimolecular phospholipid membranes'',
	Biochim.\ Biophys.\ Acta \href{https://doi.org/10.1016/0005-2736(67)90004-1}{{\bf 135}, 20} (1967)
	
\bibitem{Rosen}
	D.~Rosen and A.~M.~Sutton,
	``The effects of a direct current potential bias on the electrical properties of bimolecular lipid membranes'',
	Biochim.\ Biophys.\ Acta \href{https://doi.org/10.1016/0005-2736(68)90101-6}{{\bf 163}, 226} (1968)
	
\bibitem{White}
	S.~H.~White,
	``A Study of Lipid Bilayer Membrane Stability Using Precise Measurements of Specific Capacitance'',
	Biophys.\ J.\ \href{https://doi.org/10.1016/S0006-3495(70)86360-3}{{\bf 10}, 1127} (1970)
	
\bibitem{Wobschall}
	D.~Wobschall,
	``Voltage dependence of bilayer membrane capacitance'',
	J.\ Colloid Interface Sci.\ \href{https://doi.org/10.1016/0021-9797(72)90351-7}{{\bf 40}, 417} (1972)
	
\bibitem{Crowley}
	J.~M.~Crowley,
	``Electrical Breakdown of Bimolecular Lipid Membranes as an Electromechanical Instability'',
	Biophys.\ J.\ \href{https://dx.doi.org/10.1016\%2FS0006-3495(73)86017-5}{{\bf 13}, 711} (1973)
	
\bibitem{Feldman1986a}
	V.~J.~Feldman, M.~B.~Partenskii, and M.~M.~Vorob'ev,
	``Density functional approach to the metal-solid electrolyte interface: Electron relaxation effect, equilibrium electrical properties and bilayer instability problem'',
	Electrochim.\ Acta \href{https://doi.org/10.1016/0013-4686(86)80081-0}{{\bf 31}, 291} (1986)
	
\bibitem{Feldman1986b}
	V.~J.~Feldman, M.~B.~Partenskii, and M.~M.~Vorob'ev,
	``Surface electron screening theory and its applications to metal-electrolyte interfaces'',
	Prog.\ Surface Sci.\ \href{https://doi.org/10.1016/0079-6816(86)90010-9}{{\bf 23}, 3}  (1986)
	
\bibitem{Feldman1987}
	V.~J.~Feldman, M.~B.~Partenskii, and A.~A.~Kornyshev,
	``On the non-linear response to charging of a relaxing capacitor'',
	J.\ Electroanal.\ Chem.\ \href{https://doi.org/10.1016/0022-0728(87)80303-0}{{\bf 237}, 1} (1987)
	
\bibitem{Kornyshev}
	A.~A.~Kornyshev,
	``Metal electrons in the double layer theory'',
	Electrochim.\ Acta \href{https://doi.org/10.1016/0013-4686(89)85070-4}{{\bf 34}, 1829} (1989)
	
\bibitem{Partenskii1996}
	M.~B.~Partenskii,  V.~L.~Dorman, and P.~C.~Jordan,
	``The question of negative capacitance and its relation to instabilities and phase transitions at electrified interfaces'',
	Int.\ Rev.\ Phys.\ Chem.\ \href{https://doi.org/10.1080/01442359609353179}{{\bf 15}, 153} (1996)
	
\bibitem{Nieminen}
	H.~Nieminen, V.~Ermolov, K.~Nybergh, S.~Silanto, and T.~Ryh\"anen,
	``Microelectromechanical capacitors for RF applications'',
	J.\ Micromech.\ Microeng.\ \href{https://doi.org/10.1088/0960-1317/12/2/312}{{\bf 12}, 177} (2002)
	
\bibitem{Pershin}
	Y.~V.~Pershin and M.~Di Ventra,
	``Memory effects in complex materials and nanoscale systems'',
	Adv.\ Phys.\ \href{https://doi.org/10.1080/00018732.2010.544961}{{\bf 60}, 145} (2011)
	
\bibitem{Partenskii2001}
	M.~B.~Partenskii and P.~C.~Jordan,
	``Electroelastic Instabilities in Double Layers and Membranes'',
	in {\it Liquid Interfaces in Chemical, Biological and Pharmaceutical Applications}, Surfactant Science series {\bf 95}, ed.\ A.~G.~Volkov
	(New York: Marcel Dekker, 2001), pp.\ 51--82
	
\bibitem{Partenskii2002}
	M.~B.~Partenskii and P.~C.~Jordan,
	``Membrane deformation and the elastic energy of insertion: Perturbation of membrane elastic constants due to peptide insertion'',
	J.\ Chem.\ Phys.\ \href{https://doi.org/10.1063/1.1519840}{{\bf 117}, 10768} (2002)

\bibitem{Partenskii2005}
	M.~B.~Partenskii and P.~C.~Jordan,
	``Negative capacitance and instability at electrified interfaces: Lessons from the study of membrane capacitors'',
	Condens.\ Matter Phys.\ \href{https://doi.org/10.5488/CMP.8.2.397}{{\bf 8}, 397} (2005)
	[arXiv:physics/0412183 [physics.chem-ph]]
	
\bibitem{Partenskii2009}
	M.~B.~Partenskii and P.~C.~Jordan,
	```Squishy capacitor' model for electrical double layers and the stability of charged interfaces'',
	Phys.\ Rev.\ E \href{https://doi.org/10.1103/PhysRevE.80.011112}{{\bf 80}, 011112} (2009)
	
\bibitem{Partenskii2011}
	M.~B.~Partenskii and P.~C.~Jordan,
	``Relaxing gap capacitor models of electrified interfaces'',
	Am.\ J.\ Phys.\ \href{https://doi.org/10.1119/1.3490647}{{\bf 79}, 103} (2011)

\bibitem{AVK}
	A.~A.~Andronov, A.~A.~Vitt and S.~\`E.~Kha\u{\i}kin,
	{\it Theory of Oscillators}, ed.\ W.~Fishwick,
	(Mineola, NY: Dover, 1987 [1966]), p.\ 200
	
\bibitem{LeC1}
	P.~Le Corbeiller,
	``The non-linear theory of the maintenance of oscillations,''
	J.\ Inst.\ Electr.\ Eng.\ {\bf 79}, 361 (1936),
	reprinted in Proc.\ Inst.\ Electr.\ Eng.\ \href{https://doi.org/10.1049/PWS.1936.0030}{{\bf 11}, 292} (1936)
	
\bibitem{LeC2}
	P.~Le Corbeiller,
	``Theory of prime movers'',
	in {\it Non-Linear Mechanics}, eds.\ K.~O.~Friedrichs, P.~Le Corbeiller, N.~Levinson and J.~J.~Stoker,
	(Providence: Brown U., 1943), pp.\ 2.1--2.18
	
\bibitem{SO}
	A.~Jenkins,
	``Self-oscillation'',
	Phys.\ Rep.\ \href{https://doi.org/10.1016/j.physrep.2012.10.007}{{\bf 525}, 167} (2013)
	[arXiv:1109.6640 [physics.class-ph]]	

\bibitem{vdP1}
	B.~van der Pol,
	``On `relaxation-oscillations,' \!\!''
	Philos.\ Mag.\ (ser.\ 7) \href{https://doi.org/10.1080/14786442608564127}{{\bf 2}, 978} (1926)
	
\bibitem{vdP2}
	B.~van der Pol,
	``Forced Oscillations in a Circuit with non-linear Resistance. (Reception with reactive Triode.)'',
	in {\it Selected Papers on Mathematical Trends in Control Theory},
	eds.\ R.~Bellmann and R.~Kalaba, (New York: Dover, 1964), pp.\ 124--140.
This is a reprinting of
	Philos.\ Mag.\ (ser.\ 7) \href{https://doi.org/10.1080/14786440108564176}{{\bf 3}, 65} (1927)
	
\bibitem{active-brownian}
	W.~Ebeling, F.~Schweitzer and B.~Tilch,
	``Active Brownian particles with energy depots modeling animal mobility'',
	BioSystems \href{https://doi.org/10.1016/S0303-2647(98)00027-6}{{\bf 49}, 17} (1999)
	
\bibitem{self-propelled}
	C.~Bechinger, R.~Di Leonardo, H.~L\"owen, C.~Reichhardt, G.~Volpe, and G.~Volpe,
	``Active particles in complex and crowded environments'',
	Rev.\ Mod.\ Phys.\ \href{https://doi.org/10.1103/RevModPhys.88.045006}{{\bf 88}, 045006} (2016)
	[arXiv:1602.00081 [cond-mat.soft]]
	
\bibitem{Seifert}
	U.~Seifert,
	``Stochastic thermodynamics, fluctuation theorems and molecular machines'',
	Rep.\ Prog.\ Phys.\
	\href{https://doi.org/10.1088/0034-4885/75/12/126001}{{\bf 75}, 126001} (2012)
	[arXiv:1205.4176 [cond-mat.stat-mech]]
	
\bibitem{Minorsky}
	N.~Minorsky,
	``Self-excited Oscillations in Dynamical Systems Possessing Retarded Action,''
	in {\it Selected Papers on Mathematical Trends in Control Theory},
	eds.\ R.~Bellmann and R.~Kalaba, (New York: Dover, 1964), pp.\ 141--149.
This is a reprinting of
	J.\ Appl.\ Mech.\ \href{https://doi.org/10.1115/1.4009185}{{\bf 9}, A65} (1942).

\bibitem{bio-oscillators}
	B.~Nov\'ak and J.~J.~Tyson,
	``Design principles of biochemical oscillators'',
	Nat.\ Rev.\ Mol.\ Cell.\ Biol.\ \href{https://doi.org/10.1038/nrm2530}{{\bf 9}, 981} (2008)
	
\bibitem{Hanggi}
	P.~H\"anggi and F.~Marchesoni,
	``Artificial Brownian motors: Controlling transport on the nanoscale'',
	Rev.\ Mod.\ Phys.\
	\href{https://doi.org/10.1103/RevModPhys.81.387}{{\bf 81}, 387} (2009)
	[arXiv:0807.1283 [cond-mat.stat-mech]]

\bibitem{emf}
	R.~N.~Varney and L.~H.~Fisher,
	``Electromotive force: Volta's forgotten concept'',
	Am.\ J.\ Phys.\ \href{https://doi.org/10.1119/1.12115}{{\bf 48}, 405} (1980)
	
\bibitem{mercury}
See
	S.-W.~Lin, J.~Keizer, P.~A.~Rock, and H.~Stenschke,
	``On the Mechanism of Oscillations in the `Beating Mercury Heart'",
	Proc.\ Natl.\ Acad.\ Sci.\ U.S.A.\ \href{https://doi.org/10.1073/pnas.71.11.4477}{{\bf 71}, 4477} (1974),
and references therein.

\bibitem{shuttle1}
	L.~Y.~Gorelik, A.~Isacsson, M.~V.~Voinova, B.~Kasemo, R.~I.~Shekhter and M.~Jonson,
	``Shuttle Mechanism for Charge Transfer in Coulomb Blockade'',
	Phys.\ Rev.\ Lett.\ \href{https://doi.org/10.1103/PhysRevLett.80.4526}{{\bf 80}, 4526} (1998)
	[arXiv:cond-mat/9711196]
	
\bibitem{shuttle2}
	C.~W.~W\"achtler, P.~Strasberg, S.~H.~L.~Klapp, G.~Schaller and C.~Jarzynski,
	``Stochastic thermodynamics of self-oscillations: the electron shuttle'',
	New J.\ Phys.\ \href{https://doi.org/10.1088/1367-2630/ab2727}{{\bf 21}, 073009} (2019)
	[arXiv:1902.08174 [cond-mat.stat-mech]]

\bibitem{solarcells}
	R.~Alicki, D.~Gelbwaser-Klimovsky and K.~Szczygielski,
	``Solar cell as a self-oscillating heat engine'',
	J.\ Phys.\ A: Math.\ Theor.\
	\href{http://dx.doi.org/10.1088/1751-8113/49/1/015002}{{\bf 49}, 015002} (2016)
	[arXiv:1501.00701 [cond-mat.stat-mech]]
	
\bibitem{AGJ}
	R.~Alicki, D.~Gelbwaser-Klimovsky, and A.~Jenkins,
	``A thermodynamic cycle for the solar cell'',
	Ann.\ Phys.\ (NY)
	\href{https://doi.org/10.1016/j.aop.2017.01.003}{{\bf 378}, 71} (2017)
	[arXiv:1606.03819 [cond-mat.stat-mech]]
	
\bibitem{thermocells}
	R.~Alicki,
	``Thermoelectric generators as self-oscillating heat engines'',
	J.\ Phys.\ A: Math.\ Theor.\
	\href{http://dx.doi.org/10.1088/1751-8113/49/8/085001}{{\bf 49}, 085001} (2016)
	[arXiv:1506.00094 [quant-ph]]
	
\bibitem{fuelcells}
	R.~Alicki,
	``Unified Quantum Model of Work Generation in Thermoelectric Generators, Solar and Fuel Cells'',
	Entropy \href{http://dx.doi.org/10.3390/e18060210}{{\bf 18}, 210} (2016)
	
\bibitem{battery}
	R.~Alicki, D.~Gelbwaser-Klimovsky, A.~Jenkins, and E.~von Hauff,
	``A dynamical theory of the battery's electromotive force'',
	Phys.\ Chem.\ Chem.\ Phys.\ \href{http://doi.org/10.1039/D1CP00196E}{{\bf 23}, 9428} (2021)
	[arXiv:2010.16400 [physics.chem-ph]]
	
\bibitem{Nakanishi2012}
	H.~Nakanishi, S.~Fujiwara, K.~Takayama, I.~Kawayama, H.~Murakami, and M.~Tonouchi,
	``Imaging of a Polycrystalline Silicon Solar Cell Using a Laser Terahertz Emission Microscope'',
	Appl.\ Phys.\ Express \href{https://doi.org/10.1143/APEX.5.112301}{{\bf 5}, 112301} (2012)

\bibitem{Guzelturk2018}
	B.~Guzelturk et al.,
	``Terahertz Emission from Hybrid Perovskites Driven by Ultrafast Charge Separation and Strong Electron--Phonon Coupling'',
	Adv.\ Mater.\ \href{https://doi.org/10.1002/adma.201704737}{{\bf 30}, 1704737} (2018)

\bibitem{membrane-forces}
	A.~Anishkin, S.~H.~Loukin, J.~Teng, and C.~Kung,
	``Feeling the hidden mechanical forces in lipid bilayer is an original sense",
	Proc.\ Natl.\ Acad.\ Sci.\ U.S.A.\ \href{https://doi.org/10.1073/pnas.1313364111}{{\bf 111}, 7898} (2014)
	
\bibitem{AK}
	O.~S.~Andersen and R.~E.~Koeppe,
	``Bilayer Thickness and Membrane Protein Function: An Energetic Perspective'',
	Annu.\ Rev.\ Biophys.\ Biomol.\ Struct.\ \href{https://doi.org/10.1146/annurev.biophys.36.040306.132643}{{\bf 36}, 107} (2007)
	
\bibitem{Hodgkin-Huxley}
	A.~L.~Hodgkin and A.~F.~Huxley,
	``A quantitative description of membrane current and its application to conduction and excitation in nerve'',
	J.\ Physiol.\ \href{http://www.ncbi.nlm.nih.gov/pmc/articles/PMC1392413/}{{\bf 117}, 500} (1952)

\bibitem{FitzHugh}
	R.~FitzHugh,
	``Impulses and Physiological States in Theoretical Models of Nerve Membrane'',
	Biophys.\ J.\ \href{http://www.ncbi.nlm.nih.gov/pmc/articles/PMC1366333/}{{\bf 1}, 445} (1961)
	
\bibitem{Nagumo}
	J.~Nagumo, S.~Arimoto, and S.~Yoshizawa,
	``An Active Pulse Transmission Line Simulating Nerve Axon'',
	Proc.\ Inst.\ Radio Engrs.\ \href{http://dx.doi.org/10.1109/JRPROC.1962.288235}{{\bf 50}, 2061} (1962)
	
\bibitem{NeuronDynamics}
	B.~Ibarz, J.~M.~Casado, and M.~A.~F.~Sanju\'an,
	``Map-based models in neuronal dynamics'',
	Phys.\ Rep.\ \href{http://dx.doi.org/10.1016/j.physrep.2010.12.003}{{\bf 501}, 1} (2011)
	
\bibitem{Alfven}
	H.~Alfv\'en,
	``Double Layers and Circuits in Astrophysics'',
	IEEE Trans.\ Plasma Sci.\ \href{https://doi.org/10.1109/TPS.1986.4316626}{{\bf 14}, 779} (1986)

\bibitem{Torven}
	S.~Torv\'en and L.~Lindberg,
	``Properties of a fluctuating double layer in a magnetised plasma column'',
	J.\ Phys.\ D: Appl.\ Phys.\ \href{https://doi.org/10.1088/0022-3727/13/12/014}{{\bf 13}, 2285} (1980)
	
\bibitem{Volwerk}
	M.~Volwerk,
	``Radiation from electrostatic double layers in laboratory plasmas'',
	J.\ Phys.\ D: Appl.\ Phys.\ \href{https://doi.org/10.1088/0022-3727/26/8/007}{{\bf 26}, 1192} (1993)
	
\bibitem{Amin}
	M.~Amin,
	``Ascent of sap in plants by means of electrical double layers'',
	J.\ Biol.\ Phys.\ \href{https://doi.org/10.1007/BF01988693}{{\bf 10}, 103} (1982)
	
\bibitem{Zimmermann}
	U.~Zimmermann, H.~Schneider, L.~H.~Wegner, and A.~Haase,
	``Water ascent in tall trees: does evolution of land plants rely on a highly metastable state?'',
	New Phytol.\ \href{https://doi.org/10.1111/j.1469-8137.2004.01083.x}{{\bf 162}, 575} (2004)
	
\bibitem{Perelman}
	M.~E.~Perel'man and G.~M.~Rubinstein,
	``Ultrasound vibrations of plant cells membranes: Water lift in trees, electrical phenomena'',
	arXiv:physics/0611133 [physics.bio-ph]
	
\bibitem{Gagliano}
	M.~Gagliano,
	``Green symphonies: a call for studies on acoustic communication in plants'',
	Behav.\ Ecol.\ \href{https://dx.doi.org/10.1093\%2Fbeheco\%2Fars206}{{\bf 24}, 789} (2013)
	
	
\bibitem{Ouerdane}
	H.~Ouerdane, Y.~Apertet, C.~Goupil, and P.~Lecoeur,
	``Continuity and boundary conditions in thermodynamics: From Carnot's efficiency to efficiencies at maximum power'',
	Eur.\ Phys.\ J.-Spec.\ Top.\ \href{http://dx.doi.org/10.1140/epjst/e2015-02431-x}{{\bf 224}, 839} (2015)
	[arXiv:1411.4230 [physics.hist-ph]]
	
\bibitem{dissipation-induced}
	C.~D.~D\'iaz-Mar\'in and A.~Jenkins,
	``A physical approach to dissipation-induced instabilities'',
	arXiv:1806.01527 [physics.class-ph]
	
	
\bibitem{Strogatz}
	S.~H.~Strogatz,
	{\it Nonlinear Dynamics and Chaos}, 2nd ed.,
	(Boulder: Westview Press, 2014), sec.\ 8.2
	
\bibitem{Groszkowski}
	J.~Groszkowski,
	{\it Frequency of Self-Oscillations},
	(Oxford: Pergamon Press, 1964), sec.\ 9.4
	

\bibitem{Gouy}
	S.~L.~Carnie and G.~M.~Torrie,
	``The Statistical Mechanics of the Electrical Double Layer'',f
	Adv.\ Chem.\ Phys.\ \href{https://doi.org/10.1002/9780470142806.ch2}{{\bf 56}, 141} (1984)
	
\bibitem{powerspec}
	M.~C.~Gutzwiller,
	{\it Chaos in Classical and Quantum Mechanics}
	(New York: Springer, 1990)
	
\bibitem{heart}
	H.~Degn, A.~V.~Holden, and L.~F.~Olsen (eds.),
	\href{http://dx.doi.org/10.1007/978-1-4757-9631-5}{\it Chaos in Biological Systems},
	Nato Science Series A {\bf 138}, (1987)
	
	
\bibitem{cyclotron}
	E.~O.~Lawrence and M.~S.~Livingston,
	``The Production of High Speed Light Ions Without the Use of High Voltages'',
	Phys.\ Rev.\ \href{https://doi.org/10.1103/PhysRev.40.19}{{\bf 40}, 19} (1932)

\bibitem{pumps}
	C.~E.~Brennen,
	\href{http://brennen.caltech.edu/INTPump/pumbook.pdf}{\it Hydrodynamics of Pumps}
	(Cambridge: Cambridge U.~P., 2011), ch.\ 2
	
\bibitem{Bryant}
	D.~A.~Bryant, R.~Bingham, and U.~de Angelis,
	``Double Layers Are Not Particle Accelerators'',
	Phys.\ Rev.\ Lett.\ \href{https://doi.org/10.1103/PhysRevLett.68.37}{{\bf 68}, 37} (1992)
	
\bibitem{Compton}
	R.~G.~Compton, J.~C.~Eklund, and F.~Marken,
	``Sonoelectrochemical Processes: A Review'',
	 Electroanalysis \href{https://doi.org/10.1002/elan.1140090702}{{\bf 9}, 509} (1997)
	
\bibitem{sonoelectrochem}
	B.~G.~Pollet and M.~Ashokkumar,
	``Short Introduction to Sonoelectrochemistry'',
	in {\it Introduction to Ultrasound, Sonochemistry and Sonoelectrochemistry}, SpringerBriefs in Molecular Science
	(Cham: Springer, 2019), \href{https://doi.org/10.1007/978-3-030-25862-7_2}{pp.\ 21--39}
	
\bibitem{Eisenberg}
	R.~S.~Eisenberg,
	``Computing the field in proteins and channels'',
	J.\ Membr.\ Biol.\ \href{https://doi.org/10.1007/s002329900026}{{\bf 150}, 1} (1996)
	[arXiv:1009.2857 [q-bio.BM]]
	
	
\bibitem{moduli}
	L.~Picas, F.~Rico, and S.~Scheuring,
	``Direct Measurement of the Mechanical Properties of Lipid Phases in Supported Bilayers'',
	Biophys.\ J.\ \href{http://dx.doi.org/10.1016/j.bpj.2011.11.4001}{{\bf 102}, L01} (2012)
	
\bibitem{Cross}
	M.~Cross and H.~Greenside,
	{\it Pattern Formation and Dynamics in Nonequilibrium Systems},
	(Cambridge: Cambridge U.~P., 2009), ch.\ 3
	
\bibitem{Davydov}
	A.~S.~Davydov,
	{\it Quantum Mechanics}, 2nd ed., ed.\ D.~ter Haar,
	(Oxford: Pergamon, 1965), sec.\ 55
	
\bibitem{Zitter}
On {\it Zitterbewegung} in the relativistic Dirac equation, see
	J.~D.~Bjorken and S.~D.~Drell,
	{\it Relativistic Quantum Mechanics}
	(New York: McGraw-Hill, 1964), chs.\ 3--4,
and references therein.  The modern consensus is that this is not a physical effect because the Dirac equation for the electron should be interpreted as the equation of motion for a classical field, rather than as a quantum-mechanical wave equation.

\bibitem{Mexican-wave}
	I.~Farkas, D.~Helbing, and T.~Vicsek,
	``Mexican waves in an excitable medium'',
	Nature \href{https://doi.org/10.1038/419131a}{{\bf 419}, 131} (2002)

\bibitem{Izhikevich}
	E.~M.~Izhikevich,
	{\it Dynamical Systems in Neuroscience: The Geometry of Excitability and Bursting},
	(Cambridge, MA: MIT Press, 2007), ch.\ 2
	
\bibitem{Siegel-axon}
	A.~Siegel and H.~N.~Sapru,
	{\it Essential Neuroscience}, 3rd ed.,
	(Philadelphia: Lippincott Williams \& Wilkins, 2014), ch.\ 5
	
\bibitem{Siegel-v}
	A.~Siegel and H.~N.~Sapru,
	{\it Essential Neuroscience}, 3rd ed.,
	(Philadelphia: Lippincott Williams \& Wilkins, 2014), ch.\ 14
	
\bibitem{Smits}
	M.~Smits, A.~Ghosh, J.~Bredenbeck, S.~Yamamoto, M.~M\"uller, and M.~Bonn,
	``Ultrafast energy flow in model biological membranes'',
	New J.\ Phys.\ \href{https://doi.org/10.1088/1367-2630/9/10/390}{{\bf 9}, 390} (2007)
	
	
\bibitem{Abbott}
	L.~F.~Abbott,
	``Lapicque's introduction of the integrate-and-fire model neuron (1907)'',
	Brain Res.\ Bull.\ \href{https://doi.org/10.1016/S0361-9230(99)00161-6}{{\bf 50}, 303} (1999)
	
\bibitem{QT}
	R.~Alicki and R.~Kosloff,
	``Introduction to quantum thermodynamics: History and prospects'',
	in {\it Thermodynamics in the Quantum Regime}, eds.\ F.~Binder et al.,
	(Cham: Springer, 2019), \href{https://doi.org/10.1007/978-3-319-99046-0_1}{pp.\ 1--33}
	[arXiv:1801.08314 [quant-ph]]
	
\bibitem{tribo}
	R.~Alicki and A.~Jenkins,
	``Quantum theory of triboelectricity'',
	Phys.\ Rev.\ Lett.\ \href{https://doi.org/10.1103/PhysRevLett.125.186101}{{\bf 125}, 186101} (2020)
	[arXiv:1904.11997 [cond-mat.mes-hall]]
	
\bibitem{Prigogine}
	G.~Nicolis and I.~Prigogine,
	{\it Self-Organization in Nonequilibrium Systems}
	(New York: Wiley, 1977)
	
\bibitem{Haken}
	H.~Haken,
	{\it Synergetics: An Introduction},
	(Berlin: Springer-Verlag, 1977)
	
\bibitem{Anderson}
	P.~W.~Anderson and D.~L.~Stein,
	``Broken Symmetry, Emergent Properties, Dissipative Structures, Life: Are They Related?"
	in {\it Self-Organizing Systems: The Emergence of Order}, ed.\ F.~E.~Yates, 
	(New York: Plenum Press, 1987), \href{https://doi.org/10.1007/978-1-4613-0883-6_24}{pp.\ 445--457}
	
\bibitem{Onsager}
	L.~Onsager,
	``Reciprocal Relations in Irreversible Processes I'',
	Phys.\ Rev.\ \href{http://dx.doi.org/10.1103/PhysRev.37.405}{{\bf 37}, 405} (1931);
	``Reciprocal Relations in Irreversible Processes II'',
	Phys. Rev.\ \href{http://dx.doi.org/10.1103/PhysRev.38.2265}{{\bf 38}, 2265} (1931)
	
\bibitem{Strasberg}
	P.~Strasberg, C.~W.~W\"achtler, and G.~Schaller,
	``Autonomous implementation of thermodynamic cycles at the nanoscale'',
	Phys.\ Rev.\ Lett.\ \href{https://doi.org/10.1103/PhysRevLett.126.180605}{{\bf 126}, 180605} (2021)
	[arXiv:2101.05027 [quant-ph]]
	
	
\bibitem{HH}
	P.~Horowitz and W.~Hill,
	{\it The Art of Electronics}, 3rd ed.\
	(Cambridge: Cambridge U.~P., 2015), p.\ 71
	
\bibitem{Ramaswamy}
	S.~Ramaswamy,
	``The Mechanics and Statistics of Active Matter'',
	Annu.\ Rev.\ Condens.\ Matter Phys.\ \href{https://doi.org/10.1146/annurev-conmatphys-070909-104101}{{\bf 1}, 323} (2010)
	[arXiv:1004.1933 [cond-mat.soft]]
	
\bibitem{autorotation}
	H.~J. Lugt,
	``Autorotation'',
	Ann.\ Rev.\ Fluid Mech.\ \href{https://doi.org/10.1146/annurev.fl.15.010183.001011}{{\bf 15}, 123} (1983)
	
\bibitem{rotors}
	S.~Seah, S.~Nimmrichter and V.~Scarani,
	``Work production of quantum rotor engines'',
	New J.\ Phys.\ \href{https://doi.org/10.1088/1367-2630/aab704}{{\bf 20}, 043045} (2018)
	[arXiv:1801.02820 [quant-ph]]
	
\end{thebibliography}
\end{document}